\documentclass[twocolumn, superscriptaddress, amsmath,amssymb, aps, prl, floatfix]{revtex4-1}
\preprint{LIGO-P2500751}

\usepackage{comment}
\usepackage{graphicx}
\usepackage{dcolumn}
\usepackage{bm}
\usepackage{physics}
\usepackage{mathtools}
\usepackage{tensor}
\usepackage{xcolor}
\usepackage{hyperref}
\usepackage{float}
\usepackage{booktabs}
\usepackage{graphicx} 
\usepackage{amsmath}
\usepackage{enumitem}
\usepackage{color}
\usepackage[normalem]{ulem}
\usepackage{lineno}

\newcommand{\red}[1]{{\color{red}#1}}
\newcommand{\gwone}{\textsc{gw}\scalebox{.8}{\textsc{150914}}}
\newcommand{\gwtwo}{\textsc{gw}\scalebox{.8}{\textsc{230814}}}
\newcommand{\gwthr}{\textsc{gw}\scalebox{.8}{\textsc{250114}}}

\begin{document}
\title{Are black hole spins truly near-zero?}
\begin{flushright}
{\small LIGO-P2500751}
\end{flushright}

\author{Vaishak Prasad}
\email[]{vbp5193@psu.edu}
\affiliation{Institute for Gravitation and the Cosmos, Department of Physics, Penn State University, University Park, Pennsylvania, 16801, USA}
\author{B. S. Sathyaprakash}
\affiliation{Institute for Gravitation and the Cosmos, Department of Physics, Penn State University, University Park, Pennsylvania, 16801, USA}
\affiliation{Department of Astronomy and Astrophysics, Penn State University, University Park, Pennsylvania, 16801, USA}

\date{\today}

\begin{abstract}
The fourth gravitational-wave transient catalog, GWTC-4.0, reports \red{153} binary black hole mergers with false-alarm rates $<1\,\mathrm{yr}^{-1}$. Chirp masses are typically measured well, with the smallest fractional uncertainty being $2\%$ at the $90\%$ credible level. Spins, on the other hand, are poorly constrained: the median of the best measured spin component of the population, the effective spin, is $\chi_{\rm eff}=0.04$, with a typical $90\%$ credible uncertainty of $\Delta\chi_{\rm eff}=0.44$. The large majority - 90\% of the observed black holes, are consistent with spin-magnitude $\chi<0.57$, and are weakly aligned with the orbits. At 90\% credibility, the peaks of the inferred posteriors for spin-magnitude are found to lie in the range 0.01-0.23.
We show that this ``near-zero spins'' conclusion may be prior-driven, and that the uniform-in-magnitude spin priors lead to under-exploration of the moderate-high spin region of the parameter space. Adopting a physically agnostic prior that is uniform in the spin-vector configuration space (spin states uniform in a unit sphere) 
yields similar constraints for $\chi_{\mbox{eff}}$, but substantially different spin-magnitude inferences than GWTC-4.0. The resulting shift in spins directly impacts tests of general relativity, constraints on near-extremal Kerr remnants, as well as astrophysics, including diagnostics of formation channels and hierarchical growth. In short, \emph{the data do not require vanishing spins—the prior does,} and heeding this is essential for robust GR tests and population inferences. 
\end{abstract}

\maketitle

\textit{Introduction}—The properties of merging black holes observed by the LIGO~\cite{AdvancedLIGOCQG2015}, Virgo~\cite{AdvancedVirgoCQG2015}, and KAGRA~\cite{KAGRA2012} (LVK) collaboration---in particular, the distributions of their masses, spins, and merger rates---encode formation pathways and evolutionary histories of compact binaries. Binaries may be assembled through isolated binary evolution or through dynamical interactions in dense environments~\cite{Mandel:2018mve, Mandel:2021smh,Mapelli:2020vca, Postnov:2014tza}; broadly, the former tends to align the component spins with the orbital angular momentum, while the latter produces more isotropic spin orientations. 

Consequently, \emph{accurate} spin inference is not only essential for discriminating formation channels \cite{Talbot:2017yur, Gerosa:2017kvu}, but also for fundamental physics: spin enters the waveform phasing and ringdown \cite{Kidder:1995zr, Berti:2009kk}, so biased estimates can mimic or mask deviations from general relativity \cite{Moore:2021ibq, Gupta:2024gun}; measurements of near-extremal spins test the Kerr bound \cite{Thorne:1974ve} and inform whether mergers of sub-extremal black holes can yield extremal remnants \cite{Kesden:2008ga}; and the high-spin tail, spin–mass correlations, and spin tilts provide astrophysical evidence for hierarchical growth through repeated mergers \cite{Kimball:2020qyd, Tagawa:2021ofj, Mahapatra:2022ngs}. In short, getting the spins right matters—for astrophysics, for tests of gravity, and for the origin and growth of black holes.

Spin effects enter the waveform through spin–orbit and spin–spin couplings that modulate the phase and amplitude at post-Newtonian (PN) orders higher than the leading point-mass terms \cite{Blanchet:2013haa}. The dominant spin contribution appears at 1.5PN, i.e., at order $(v/c)^3$ in the expansion parameter $v$, the velocity of the reduced mass $\mu$ in the system’s gravitational potential. At that order, the spin–orbit term, largely controlled by the effective spin,
\begin{equation}
\chi_{\mathrm{eff}} \equiv \frac{m_1 \chi_{1z} + m_2 \chi_{2z}}{m_1+m_2},
\label{eq:chi effective}
\end{equation}
is partially degenerate with the binary’s mass ratio, rendering $\chi_{\mathrm{eff}}$ significantly harder to measure than the chirp mass (which is exquisitely determined by the leading inspiral phasing) and, typically, harder than the mass ratio itself \cite{Baird:2012cu}.

Gravitational-wave measurements to date \cite{LIGOScientific:2025hdt, LIGOScientific:2025yae, LIGOScientific:2025slb} indicate that most systems cluster near zero effective spin, $\chi_{\mathrm{eff}}\approx 0$ \cite{LIGOScientific:2025hdt, LIGOScientific:2025pvj}. The population median is $0.04$, with minimum and maximum values $-0.03$ and $0.4$, respectively. Individual event posteriors also tend to peak close to zero spin magnitude \cite{LIGOScientific:2025pvj}. There have also been independent works in the past that attempt to quantify the spin orientations and how small the spins of observed black holes in binaries are \cite{Adamcewicz:2025phm, Banagiri:2025dxo}, with some finding evidence for negligible spins \cite{Tong:2022iws}, and others that suggest otherwise \cite{Callister:2022qwb}. While a variety of spin priors have been used in the past (see e.g. \cite{GWTC1, Callister:2022qwb}), the most-popular one involves choosing spin magnitudes to be uniformly distributed in $[0, 1]$, and directions isotropic on a unit sphere. It is also widely referred to and believed to be uninformative.

In this letter, we show that this result may be a direct consequence of the choice of Bayesian spin priors adopted in the analysis, and that our current spin inferences are typically likelihood-limited. We also show that the uniform-in-magnitude priors we currently use are not as uninformative and natural as one might expect. Furthermore, although the individual event priors are weighted out during population inference, we find evidence that they may still affect the population inference, due to the failure of reweighting.

\textit{LVK spin priors}—The LVK population analyses typically assume the distribution of the component spins to be isotropic in orientation, and uniform in $[0,1]$ in the dimensionless spin magnitude $\chi' = \vert\vec{\chi'}\vert$, i.e. ~\cite{LIGOScientific:2025pvj}:
\begin{align}
    p(\chi') =& 1, \qquad \chi'\in[0,1)  \\
    \int \chi' d\chi' =& 1
    \label{eq:magnitudeM}
\end{align}
for a vector $\chi'$ that lives in the vector space $V'$.

This choice is seemingly “uninformative” in the component spin magnitudes $\chi'_1$ and $\chi'_2$. However, we point out that this implicitly assumes that the spin vectors lie in a configuration space $V'$ endowed with a topology $S^2 \times I, I \in [0, 1]$. This is because the spin magnitude is not an independent one-dimensional variable, but rather the magnitude $\chi$ of a vector $\vec{\chi}$ that lives in a 3-dimensional spin configuration space $V$ that has a topology of a spherical volume, i.e., a 3-ball $B^3$. Thus, when combined with isotropy, we show below that the uniform in spin magnitude prior $p_{V'}(\chi')$ for $\chi'$ in $V'$ induces a highly biased, non-uniform geometry for its distribution $p_{V}(\chi')$, and for its Cartesian \textit{components} in the canonical Eucledian configuration space $V$.
\begin{figure}
    \centering
    \includegraphics[width=6cm, height=6cm]{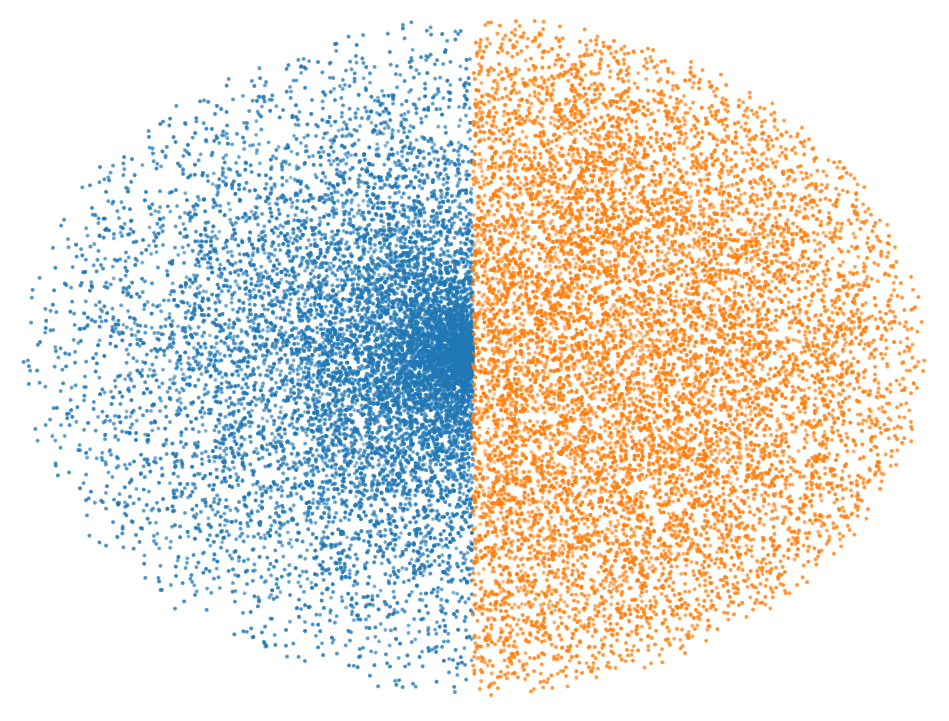}
    \caption{Distribution of the endpoints of the spin vectors. The left half contains points drawn from a uniform distribution in magnitude (blue) and an isotropic-in-direction distribution. The points in the right hemisphere (orange) are drawn uniformly in the volume, and as a consequence, are also isotropic in directions. Each plot contains about $10^5$ points.}
    \label{fig:priors3D}
\end{figure}
\begin{figure*}
    \centering
    \includegraphics[width=0.490\linewidth]{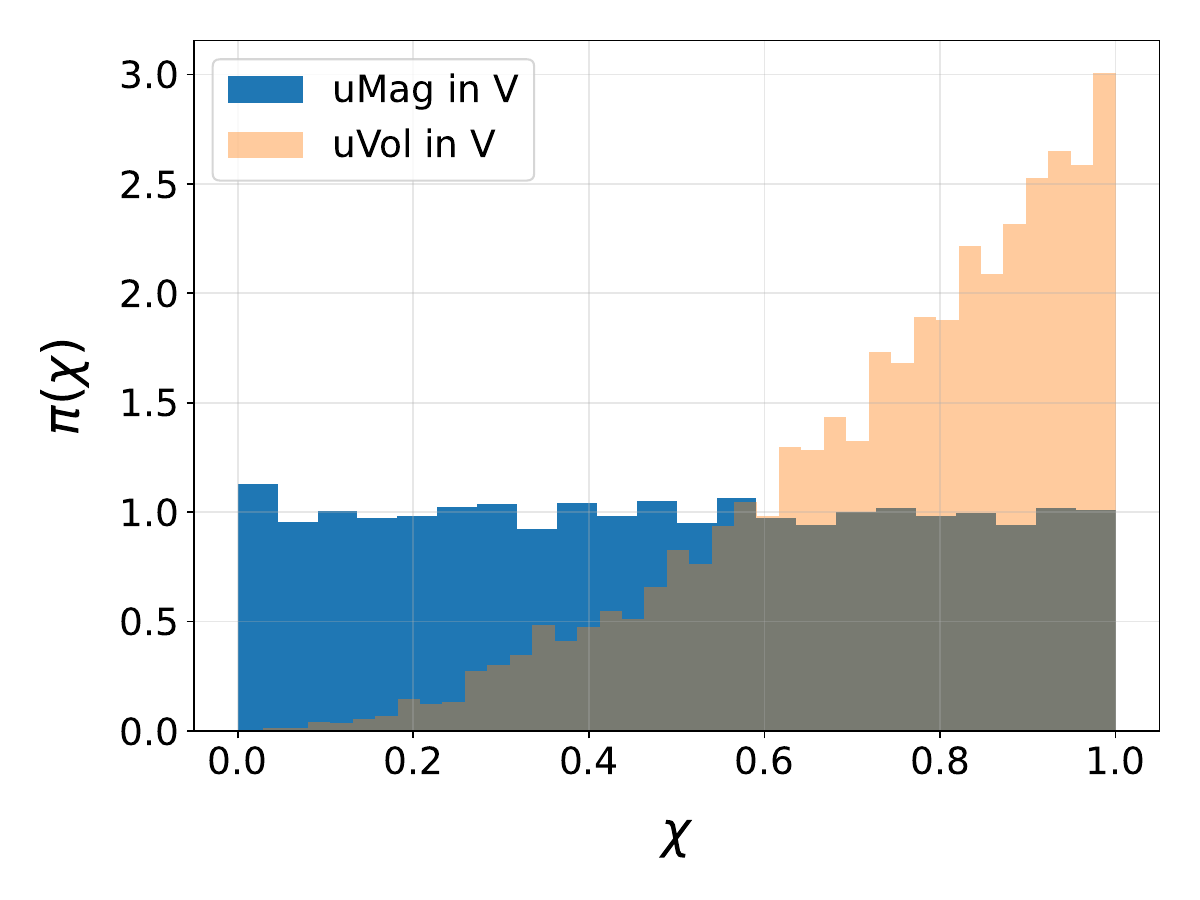}
    \includegraphics[width=0.490\linewidth]{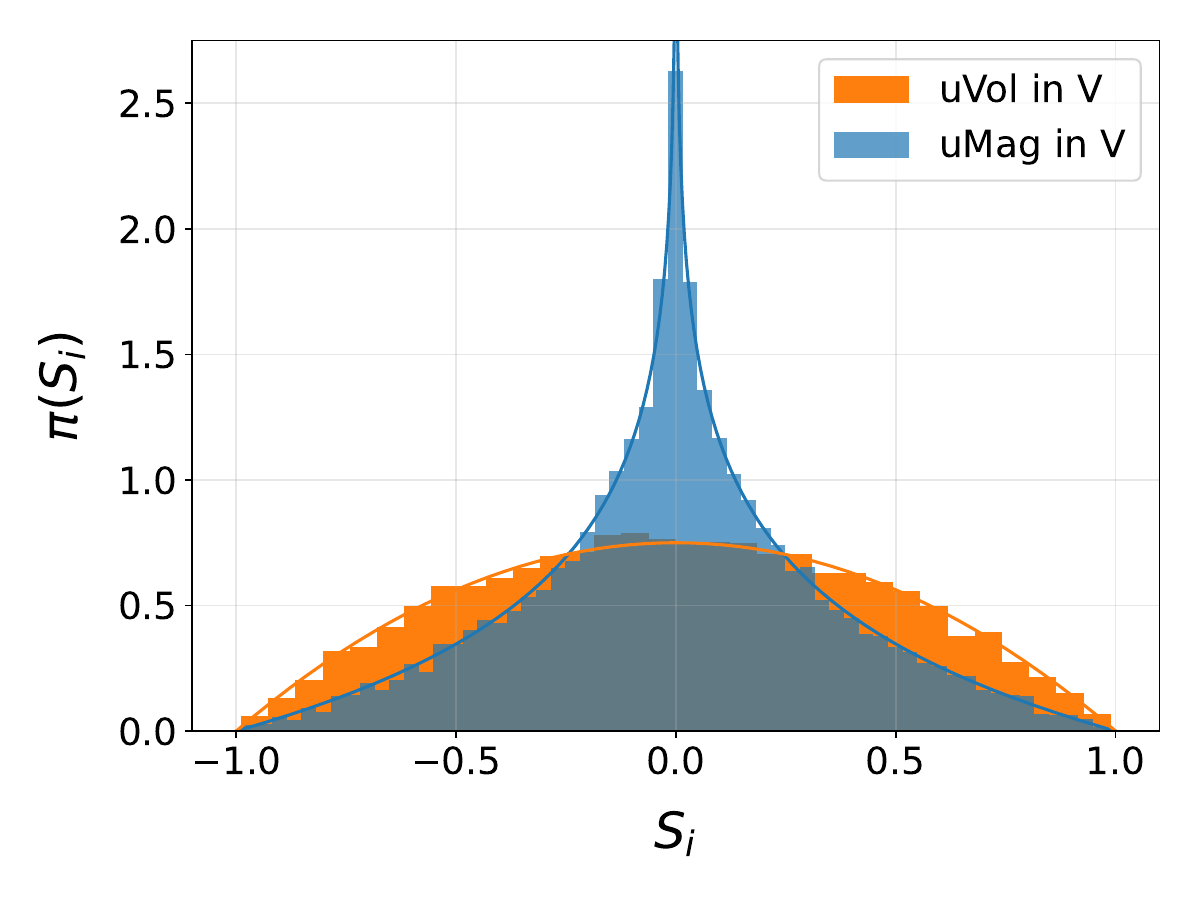}
    \caption{Marginal distribution $p'(\chi')$ of the spin magnitude $\chi'$ (blue, left panel) in the space $V'$, and that ($p(\chi)$) of  spin magnitude $\chi$ in the space $V'$ (orange, left panel). Although $\chi'$ is seemingly uninformative in $V'$, it is heavily biased towards lower spins in 3D (see Fig.~\ref{fig:priors3D}). Similarly, the seemingly biased power-law marginal distribution $p(\chi)$ is actually uniform in volume, as can be seen in \ref{fig:priors3D}. The solid orange line denotes the analytic distribution in \eqref{eq:magnitudeV}. The marginal distributions in $V$ of any of the Cartesian components $\chi_k$ are shown on the right panel. The uniform in spin-magnitude distribution is shown in blue, and the uniform in spin configuration volume is shown in orange. The solid line denotes the logarithmic distribution derived in \eqref{eq:pdfMcomponents}. In both plots, the histograms are sampled from the prior distributions for spin magnitude $\chi$ under the assumption of isotropic spin directions, from which the samples for the Cartesian components are deduced.}
    \label{fig:priors}
\end{figure*}
To see this, let us first derive the truly agnostic priors $p_V(\chi)$ for the spin vectors $\vec{\chi}$ in $V$. Given the constraint $\chi = \vert \vec{\chi} \vert <1$, it is easy to see that the spin vectors have their endpoints inside a unit round sphere $\mathcal{S}$ of radius $1$ in the spin-configuration space $V$. Let us label these endpoints by their coordinates $\chi, \theta, \phi$ in a spherical-polar coordinate system. For a truly agnostic distribution for the spin vectors, their endpoints must be uniformly distributed inside the unit sphere $\mathcal{S}$ volume. This implies the following distributions for $\chi, \theta, \phi$. The angular distributions are the same in $V$ and $V'$, and the subscript will be ignored for them:
\begin{align}
    p_V(\chi) = 3 \chi^2, \,\,\, p(\theta) = \dfrac{1}{2}sin(\theta), \,\,\, p(\phi) &= \dfrac{1}{2\pi}, \,\,\, \textit{with}  \\
    \oint_{\mathcal{S}} p(\chi)p(\theta)p(\phi) d\chi d\theta d\phi &= 1
\end{align}
We can now find what the uniform distribution $p_{V'}(\chi')$ in $V'$ means for the distribution of the points $p_V(\chi')$ in the three-dimensional spin configuration volume $V$. First note that, given the radial integration measure in $V$, for a variable $\chi'$ to have a uniform radial distribution in $V$, it must be related to the spin magnitude $\chi$ through $\chi' = \chi^3$. This implies that the corresponding distribution $p_{V}(\chi')$ for the spin magnitude $\chi'$ in the three-dimensional configuration space $V$ itself is:
\begin{align}
    p_V(\chi') \propto \dfrac{1}{\chi'^2}
\end{align}
$p_V(\chi')$ clearly exhibits a singularity at $\chi=0$. Furthermore, the marginal distribution 
$p_V(\chi_i')$ of the Cartesian components of the spins $\chi_i$ in $V$ will be logarithmic in $\chi_z$, i.e., $p'(\chi_z) \propto \ln\vert\chi_z\vert$ (see End Matter for details), with the probability being infinite at zero and also a non-differentiable cusp. However the singluarity is integrable. We show this explicitly in Figs.~\ref{fig:priors3D} and \ref{fig:priors}. A proper derivation is provided in the End Matter.


%




Consequently, even before confronting the data, drawing random points from this prior $p_V(\chi')$ concentrates the spin vectors in $V$ near the origin and thus disfavors any appreciable aligned or anti-aligned spins. This is akin to choosing, e.g., uniform priors in the distance. Consider the scenario where the cosmological homogeneity principle had not been discovered yet. In the absence of astrophysical priors for the location of the source given by its right ascension, declination, and luminosity distance $(r_a, \delta, d_{\mathcal{L}})$, we would adopt an ``agnostic" prior. As we would have no preference for the direction to the source, we would demand isotropy. For the luminosity distance, would it be appropriate for the priors to be uniform-in-distance or uniform-in-volume prior, i.e., a quadratic power-law in distance? Although Bayesian inference endows us with the freedom to choose any prior we like, and none of the priors are therefore ``incorrect", it is in our interest to adopt those priors that do not bias our perception of the universe. Especially in the absence of knowledge of the astrophysical spin distribution, it is important to address what ``agnostic" would imply.

The LVK spin priors adopt a particular type of $S^2\times R$ topology with a restricted interval, i.e. $S^2 \times I, I\in[0, 1]$, so both spaces are compact. The boundary $\partial V'$ of such a configuration space $V'$ has two disjoint elements, and can be expressed as $S^2 \times \{0\} \cup S^2 \times \{1\} $, in contrast to the boundary $\partial V$ of a space $V$ with topology $B^3$, which has only one element $S^2 \times \{1\}$. This is because, although both spaces are compact, the difference between the two topologies is that the three-ball $B^3$ can be obtained from $S^2 \times I$ by collapsing the inner boundary with topology $S^2$ to a point, where the integration measure thus vanishes. This means that in $V$, all the spin configurations with different directions, but zero magnitude, degenerate to a single state, resulting in zero probability density there. In contrast, for the case of $V'$, the possible configurations of zero-magnitude spin vectors are infinitely many and non-degenerate, and live on the lower boundary shell $S^2 \times {0}$ with topology $S^2$, where the integration measure is non-zero. Therefore, although there exists only one physical black hole spin state with spin-magnitude zero, the uniform-in-magnitude prior counts this multiple times over, assigning the same number of states to $\chi'=0$ as with any other state with non-zero spin, thus biasing the prior to the zero spin state. Thus, physically, in the absence of astrophysical information, if we are allowing for arbitrary spin configurations (i.e., including precession) and giving equal weight to every spin configuration, then the agnostic prior distribution is required to vanish at $\vec{\chi}=0$. This does not mean one cannot observe systems with zero spins as claimed in \cite{LIGOScientific:2025pvj}. On the other hand, if we are not looking for spin-precessing systems, then the canonical phase space of spin configurations will not live in $B^3$, and it can be argued that the prior probability of zero spin should be non-zero. 

Furthermore, an agnostic spin-prior distribution should also acknowledge the presence of a larger number of spin configurations at larger spin magnitudes. There are infinitely many ways to count the number of states at a spin-magnitude $\chi$, satisfying the requirement $\pi(\chi\xrightarrow{}0) \xrightarrow{}0$. They belong to a family of Beta distributions $\beta (p, b, \chi) \propto \chi^{p-1} (1-\chi)^{q -1}$ with the requirement that $p > 1$ and $q=1$ i.e. a power law in $\chi$. 

One can also examine the symmetry groups of maximally symmetric solutions of these respective topologies. In the case of $V$, each element is denoted by $\vec{\chi}$ and the reflected element by $-\vec{\chi}$. As $-\vec{\chi} \in V$, there exists continuous reflection symmetry in the group. On the other hand, in $V'$, each element of the vector space is denoted by $(\chi', \hat{\chi'})$, where $\chi' \in I$ is the spin magnitude and $\hat{\chi'} \in S^2$ the unit vector denoting the direction. It is important to note that the reflected element $(-\chi', -\hat{\chi'}) \not\in V'$, and thus the continuous reflection symmetry is broken. Thus, although the subspace $S^2$ still retains the continuous reflection symmetry, the total space does not. The natural maximal symmetry groups of these spaces are therefore different. For $V'$, it is $SO(3)\times Z_2$. The $Z_2$ exists because of the new discrete reflection symmetry about the two boundaries: $\chi' \xrightarrow{} 1-\chi'$. Whereas, in the case of $V$, it is $O(3)$, as it is naturally endowed with continuous reflection symmetry about the origin. It follows from these features that the space $V'$ of unifom-in-magnitude spins cannot be embedded in $R^3$ (but can be in $R^4$), while $V$ can be. For these reasons, it can be argued that, in the absence of additional astrophysical information, the preferred topology of spin vectors in three dimensions should be $B^3$.

Arguably, the uniform-in-magnitude prior structure can bias individual-event posteriors toward $\chi_{\mathrm{eff}}\simeq 0$. Cumulatively, this can steer the population fits toward models peaking at vanishing spin magnitude—with only occasional posteriors away from zero, and a small subset apparently supporting large spins (see Table~\ref{tab:gwtc4-summary}) when the likelihood has sufficient information. This begs a critical question: are the observed low-spin inferences a robust astrophysical signature?

\textit{Agnostic spin priors}—We propose a geometrically motivated, truly agnostic alternative that acknowledges the canonical topology of three-dimensional vectors in Euclidean space - i.e., the 3-Ball. We demand that the distribution of the spin \textit{vector} be uniform in the three-dimensional spin configuration volume. The draws from such a prior would ensure the spin vectors to be isotropic with uniform probability density inside the unit 3-ball $V$. 


As shown earlier, this uniform-in-volume distribution implies the marginal magnitude density

\begin{equation}
p_V(\chi) = 3 \chi^{2},
   \label{eq:magnitudeV}
\end{equation}

and smooth, finite Cartesian spin–component marginals $p(\chi_i) \propto (1-\chi_i^2)$, removing the cusp seen in the RHS of Fig.~\ref{fig:priors} at zero, and thus not accumulating the probability there. The prior for the individual Catesian components $S_i$ still centers on zero due to symmetry, and all spin configurations are equally probable. 

Fig.~\ref{fig:priors3D} shows the distributions of the endpoints of the vector in $V$ from the two priors. In Fig.~\ref{fig:priors}, we show the spin magnitude $\chi$ (left) and Cartesian components (right) for the two spin priors. 

Since $p_V(\chi)\propto\chi^2$, the prior appropriately acknowledges the presence of more spin-states at larger values of $\chi$, as the number of possible states scales with the area of the shell at $\chi$. 

\begin{figure}
    \centering
    \includegraphics[width=\linewidth]{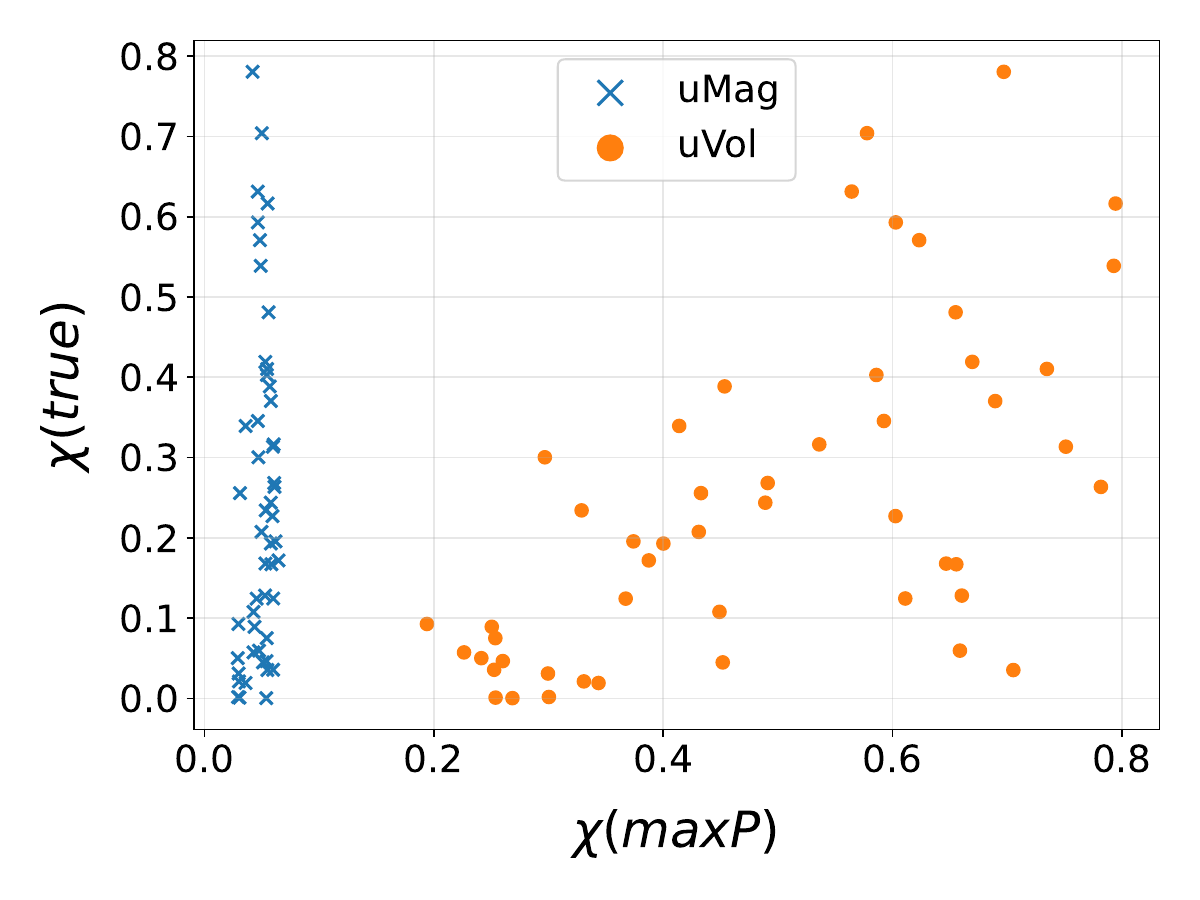}
    \caption{The most probable value of the spin magnitude vs the maximum likelihood value of the spin magnitude, from 50 draws of the spin vectors from a uniform in magnitude and isotropic in directions prior distribution 
    This clearly portrays the effect of preferential sampling near the origin (as shown in Fig.~\ref{fig:priors3D}) when the existing priors are used to infer the posterior.}
    \label{fig:vec_rec}
\end{figure}

The uniform-in-volume spin priors have been used in the past to analyze individual events \cite{lange2018rapidaccurateparameterinference, GWTC1, Vitale:2017cfs, Chatziioannou:2018wqx, DeRenzis:2022vsj}, but to the best of the authors' knowledge, the aforementioned issues have neither been pointed out nor discussed in the literature. 

To put this in perspective, we draw 50 spin vectors from a uniform distribution in magnitude and a uniform distribution in configuration space volume, and infer the posterior using Bayesian inference. We define the likelihood function as half the Euclidean distance between the input and the true vector, and estimate the posterior using both priors. We then plot the values of the spin magnitude corresponding to the maximum probability point and the maximum likelihood point. We find that using uniform-in-magnitude priors leads to a large mismatch between the most probable value and the maximum likelihood value, as shown in Fig.~\ref{fig:vec_rec}. When using a uniform-in-spin magnitude prior, the posterior almost always peaks at low values of the spin magnitude, although the best-fitting maxL value lies far from it, leading to a biased perception. One recovers similar results when the true vectors are drawn from a uniform-in-volume distribution. The results of this exercise are consistent with those in \cite{Chatziioannou:2018wqx, DeRenzis:2022vsj}, which found that uniform-in-magnitude priors hinder the unbiased recovery of non-zero / higher spin sources.

In gravitational wave astronomy, we do not directly observe spin vectors. Their information is encoded in the morphology of the observed signal, and thus, the likelihood surface will be much more complex. To fully understand the performance and compatibility of the agnostic priors for gravitational wave inference, one would have to conduct detailed injection studies, which are underway and will be reported in future work. However, we show below the results for three of the exceptional events.
\begin{figure*}
    \centering
    \includegraphics[width=0.32\linewidth]{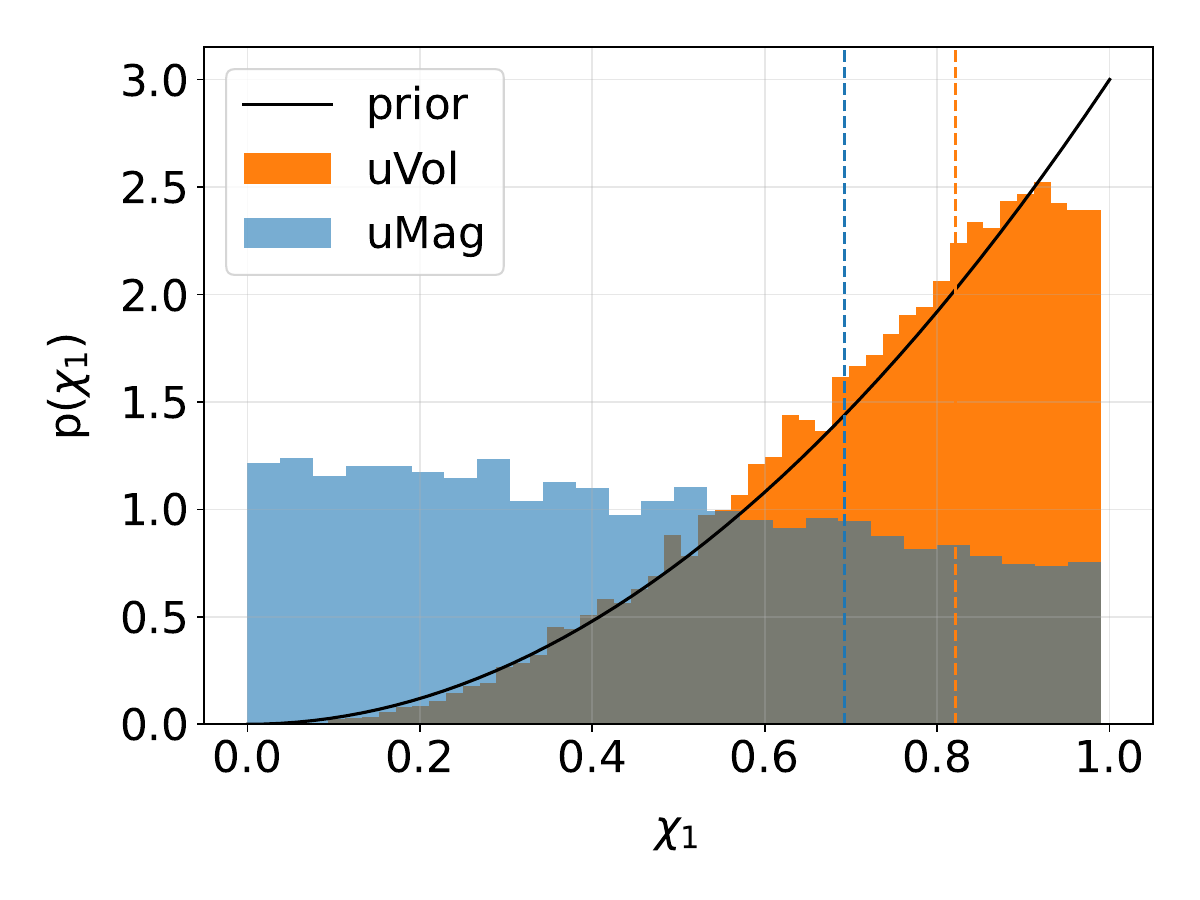}
    \includegraphics[width=0.32\linewidth]{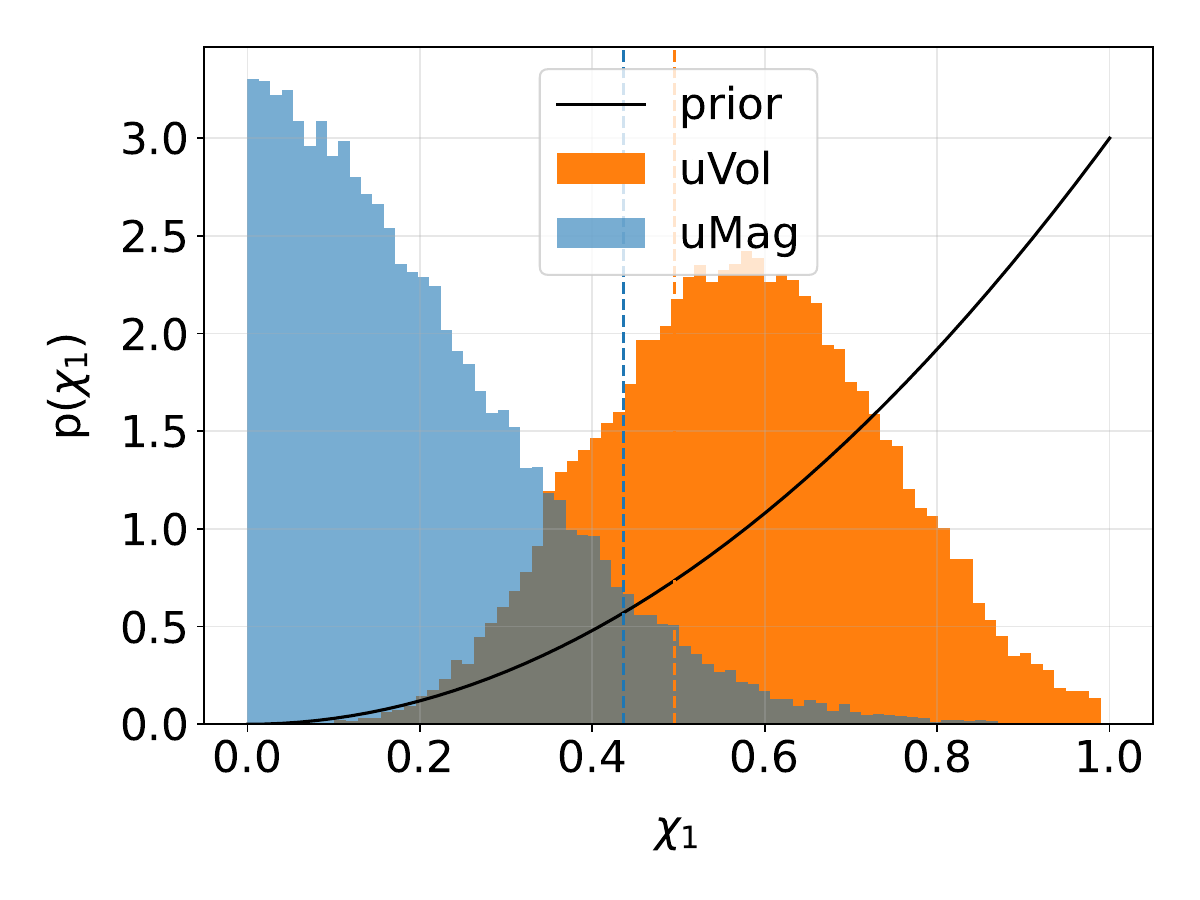}
    \includegraphics[width=0.32\linewidth]{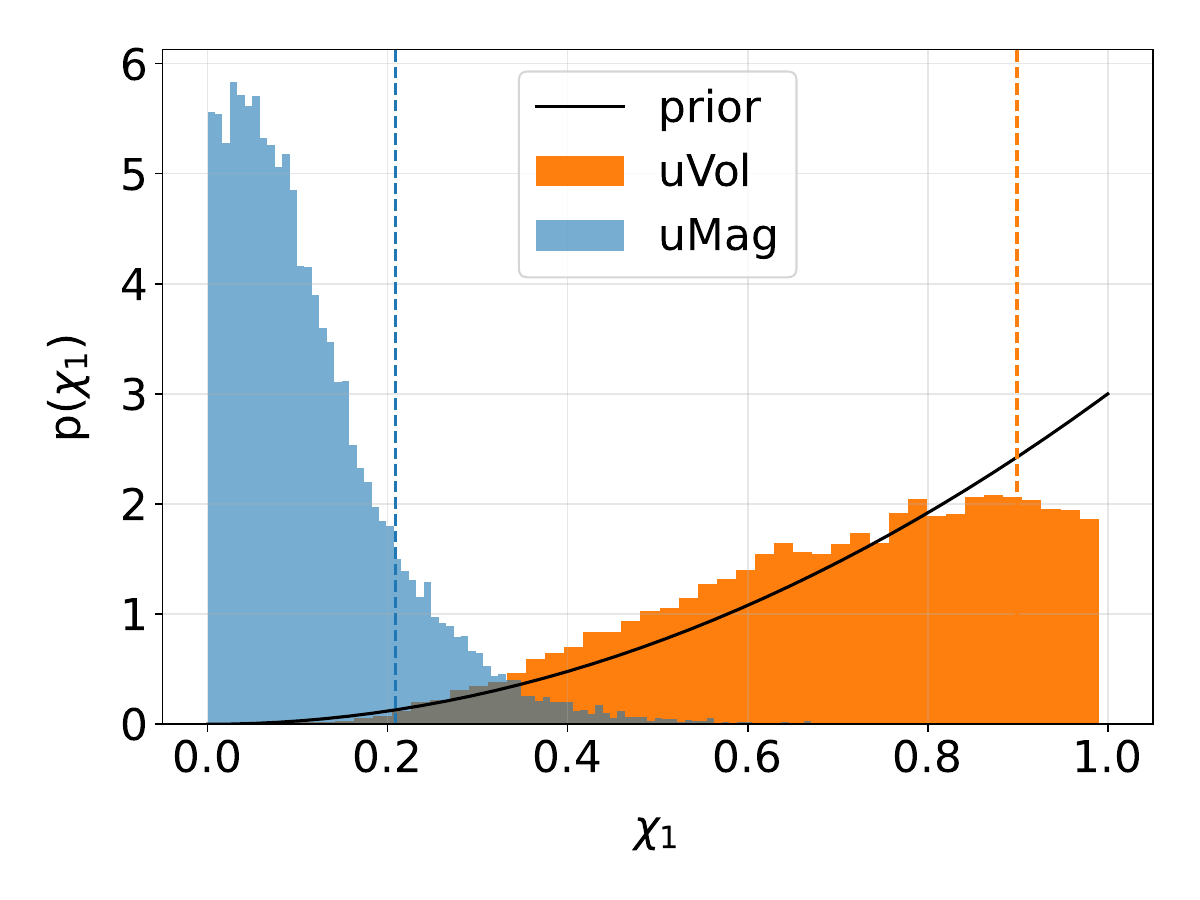}\\
    \includegraphics[width=0.32\linewidth]{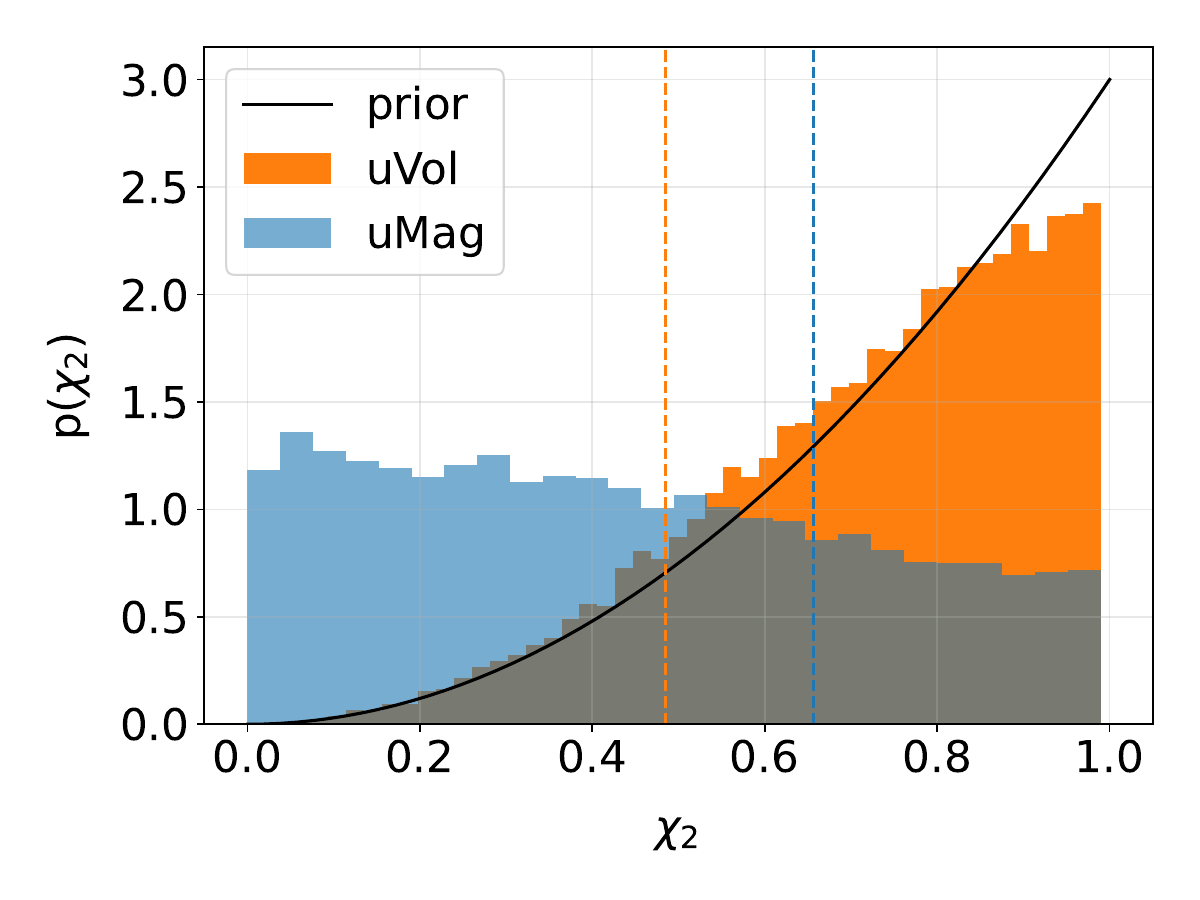}
    \includegraphics[width=0.32\linewidth]{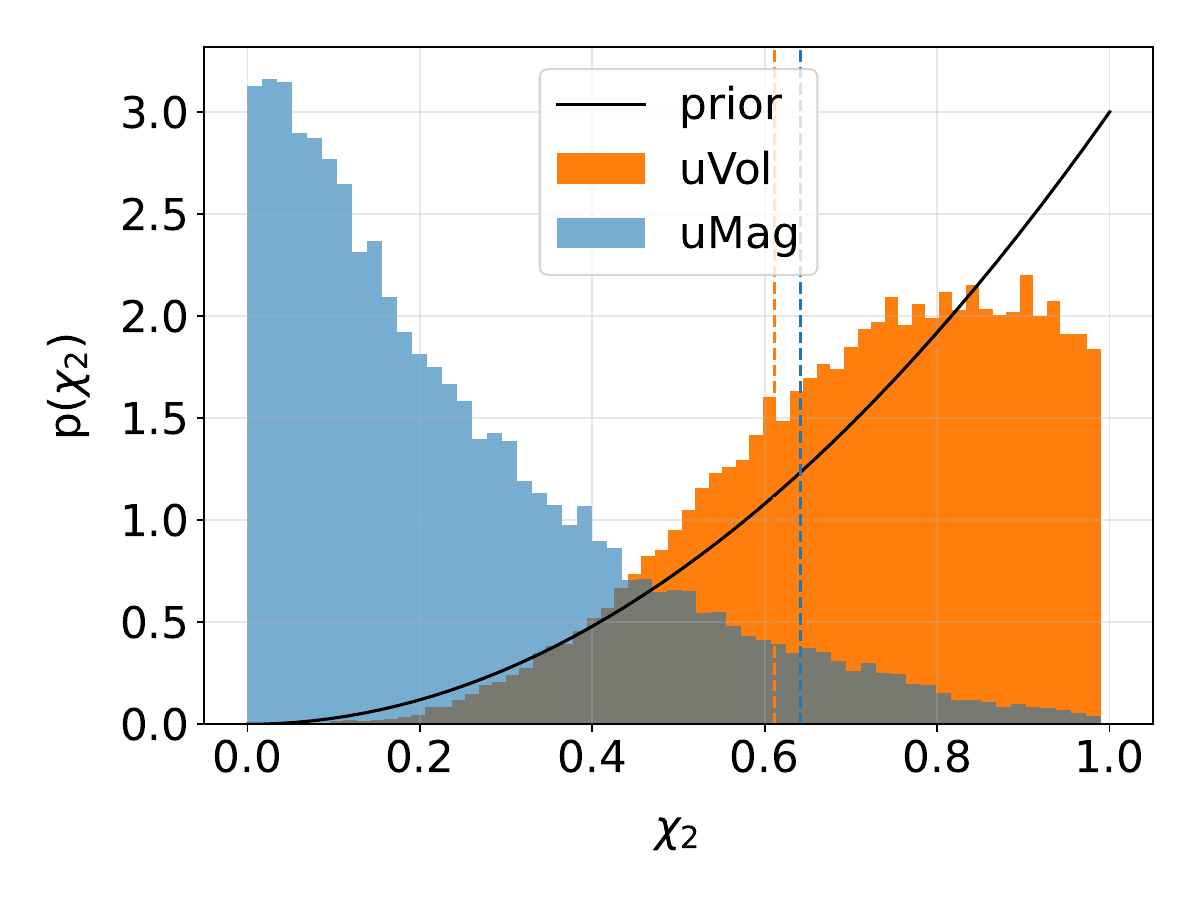}
    \includegraphics[width=0.32\linewidth]{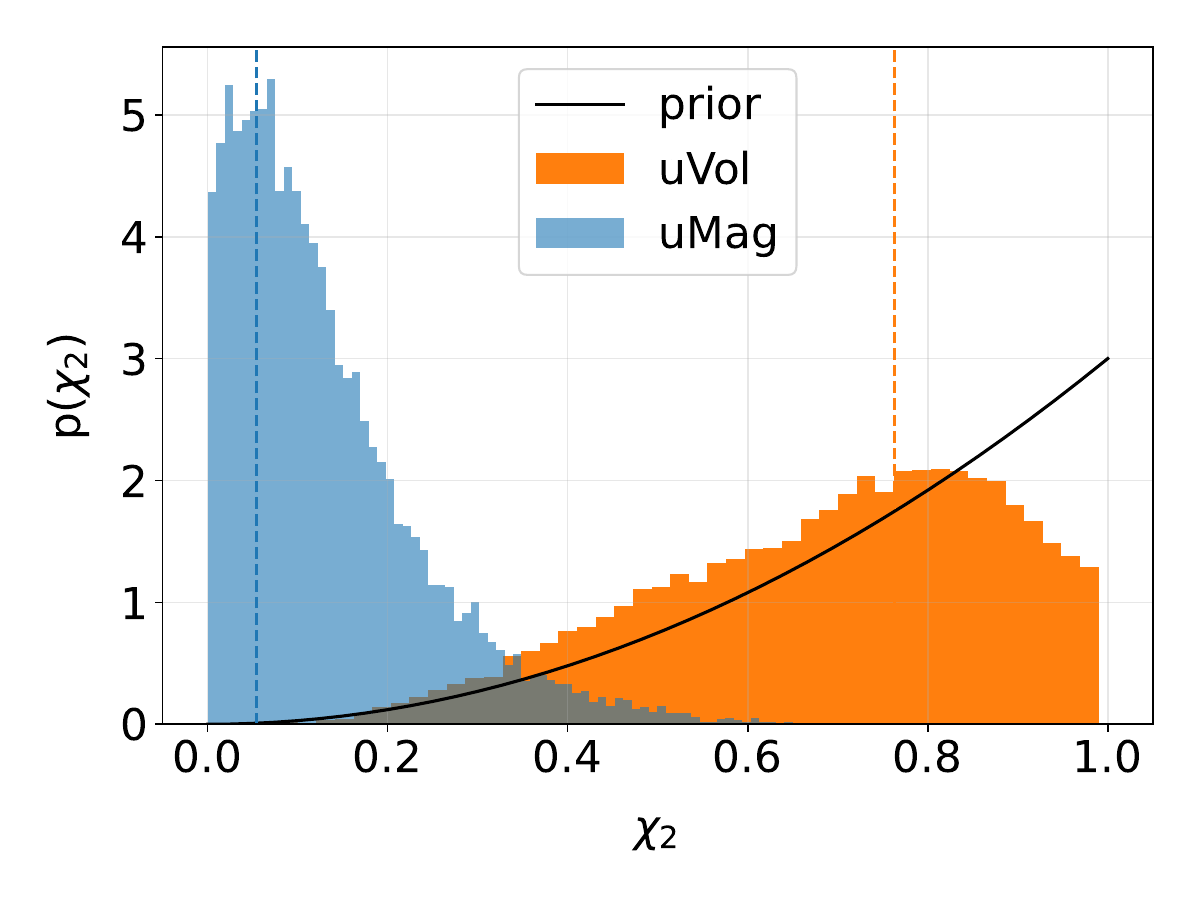}
    \caption{Marginal distributions for the spins magnitudes of GW150914 (left), GW230814 (centre) and GW250114 (right). In the case of GW150914, there is insufficient information in the likelihood, and in both cases, the prior is recovered for the posterior. One can see that the posterior distribution has better support around the region of maxL when there is sufficient information in the likelihood (GW230814 and GW250114). In the case of GW250114, the new priors also end up accumulating posteriors in an entirely different, high-spin region of the spin configuration space, where they find a marginally higher log-likelihood value.}
    \label{fig:event}
\end{figure*}
\begin{table}[h!]
\centering
\caption{Summary of spin parameters for selected gravitational-wave events, analyzed with the uniform-in-magnitude (m) and uniform-in-volume (v) priors. Although the constraints on $\chi_{eff}$ do not change, the constraints on $\chi_p$ change significantly. The last column denotes the ratio of the maximum likelihood value attained by the run using uniform-in-volume prior, to that of the uniform-in-magnitude run.}
\label{tab:const}
\renewcommand{\arraystretch}{1.5}
\begin{tabular}{l| c c c}
\hline
\textbf{Event} & $\boldsymbol{\chi_{\mathrm{eff}}}$ & $\boldsymbol{\chi_p}$ & \textbf{maxL} ratio \\
\hline
GW150914 (m) & $-0.02 ^{+0.12} _{-0.12}$ & $0.53^{+0.37} _{-0.41}$  & $1$ \\
GW150914 (v) & $-0.00 ^{+0.13} _{-0.13}$ & $0.725^{+0.21} _{-0.31}$ & $1.08$ \\
\hline
GW230814 (m) & $-0.01 ^{+0.05} _{-0.06}$ & $0.19^{+0.29} _{-0.15}$  & $1$ \\
GW230814 (v) & $0.01 ^{+0.07} _{-0.08}$  & $0.47^{+0.28} _{-0.24}$  & $1.72$ \\
\hline
GW250114 (m) & $-0.04 ^{+0.03} _{-0.04}$ & $0.12^{+0.21} _{-0.09}$ & $1$ \\
GW250114 (v) & $-0.07 ^{+0.04} _{-0.04}$ & $0.71^{+0.21} _{-0.34}$ & $2.49$ \\
\hline
\end{tabular}
\end{table}

\textit{Spins of \gwone, \gwtwo, and \gwthr}—We examine the spin distributions of three exceptional events with the new prior introduced in this Letter and compare them to results obtained with LVK priors. 

Reanalyzing \gwone, \gwtwo, and \gwthr\ under this prior, we find that the companion spins and $\chi_{\mathrm{eff}}$ shift appreciably away from zero relative to LVK-standard inferences, the median likelihood values show positive improvement consistently, by $54.3, 146.0,$ and $100.4$ percent, whereas the corresponding loglikelihood values improve by $0.03, 0.08$ and $0.14$ percent, and so do the maximum likelihood (maxL) values. The Bayes factors between the two Bayesian models, keeping the likelihood fixed, are $\sim 2.5$ for GW150914 and GW230814, and $\sim 9.5$ for GW250114 in favour of the uniform-in-magnitude model. It is to be noted, however, that Bayes factors are event-specific and cannot be used to decide the correctness of priors. A narrow, unrealistic prior closer to the maximum likelihood region can lead to high Bayes factors. 

The posteriors on the spin-magnitudes are shown in Fig.~\ref{fig:event}. For GW150914, the posteriors are almost identical to the spin priors, indicating no additional information in the likelihood functions and thus no preference for spins. The maxL value for GW230814 is very close to the one original values, but the posterior has better support around this region with the new priors. In the case of GW250114, although the recovered maxL value is only slightly larger than the original, the spin configuration corresponding to this point is very different from the original. The posteriors also have the maximum support from around this region of the parameter space. 

The constraints on the spin parameters and the increase in likelihood values are shown in Tab.~\ref{tab:const}. Across these events, a consistent pattern can be noticed. The choice of the prior does not seem to change the constraints on the beast measured spin-parameter $\chi_{eff}$. However, the value of the in-plane components or $\chi_p$ changes noticeably. This can be explained by the same feature of the uniform-in-magnitude priors -- they prefer zero spins and do not explore the high-spin, precessing region of the parameter space well.
%
%




These results show that prior geometry can materially affect both event-level and population-level inferences, motivating a re-examination of BBH spin distributions with priors that are uniform in the underlying spin \textit{geometry} rather than the spin \textit{magnitude}. 
\begin{figure}
    \centering
    \includegraphics[width=\linewidth]{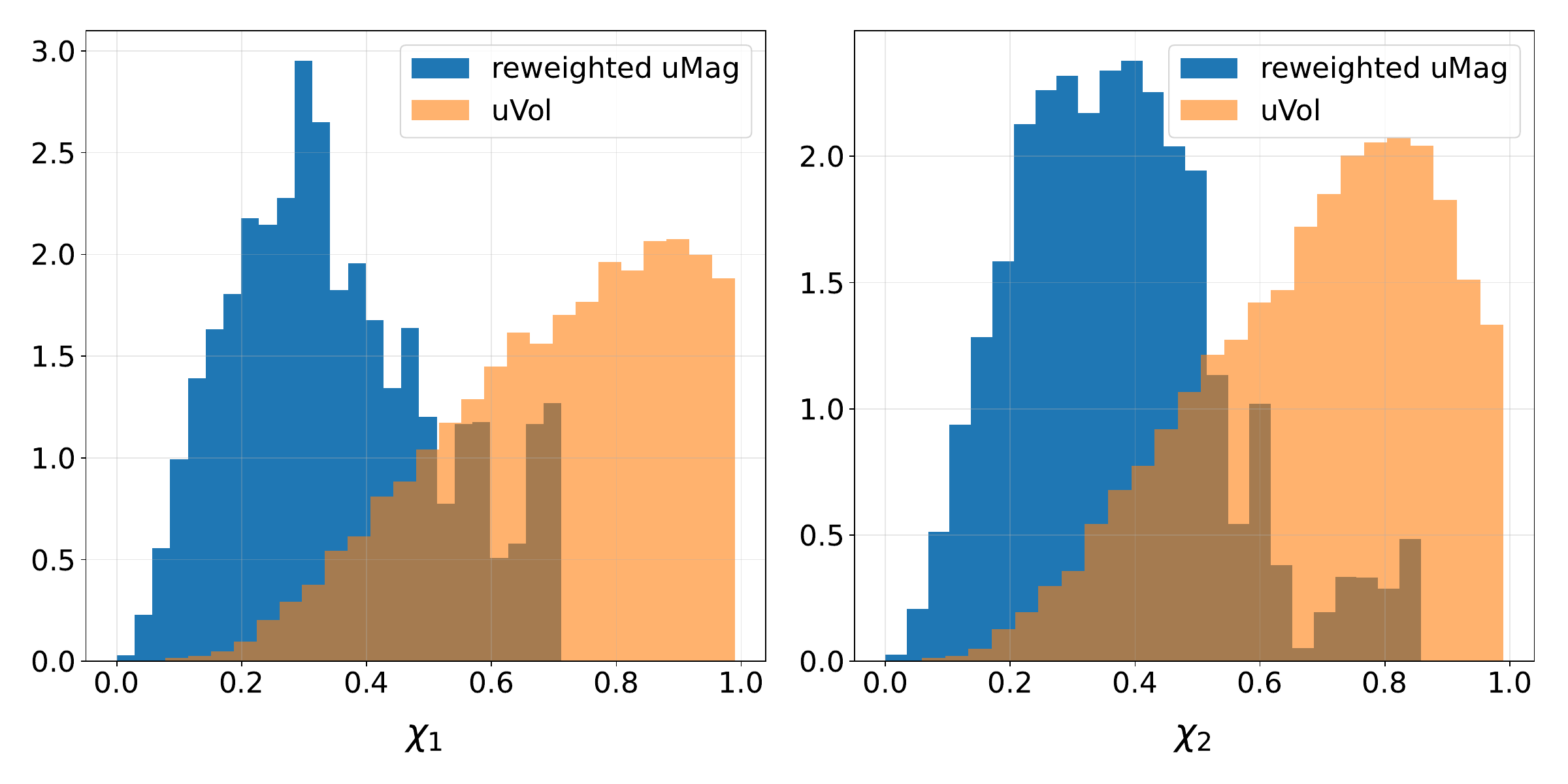}\\
    \caption{Performance of reweighting of posterior samples with uniform-in-magnitude priors to obtain the marginal posterior distribution for the spin magnitudes with uniform-in-volume priors, in the GW250114 analysis. The reweighted samples are in blue, while the samples run with uniform-in-volume priors are in orange. 
    }
    \label{fig:reweight}
\end{figure}
%


\begin{figure*}
    \centering
    \includegraphics[width=0.49\linewidth]{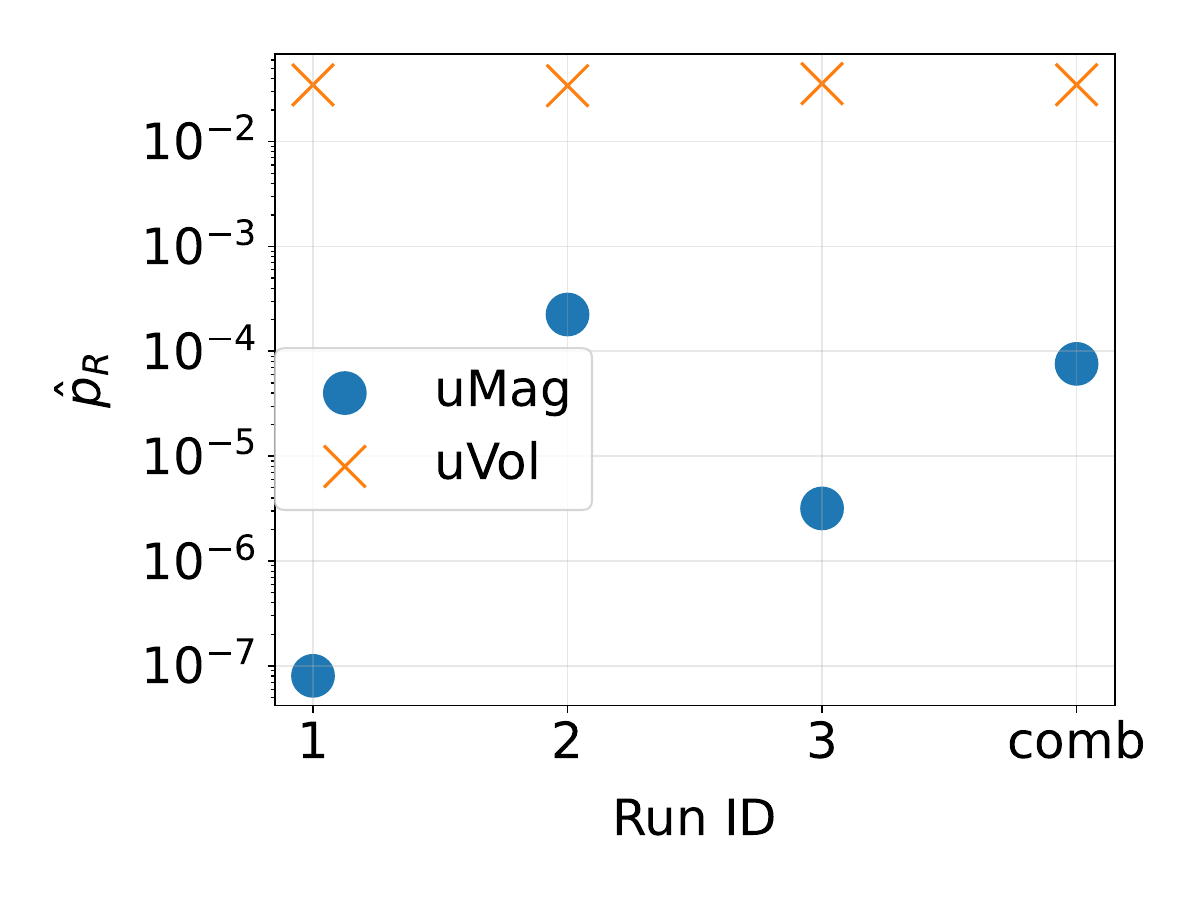}
    \includegraphics[width=0.49\linewidth]{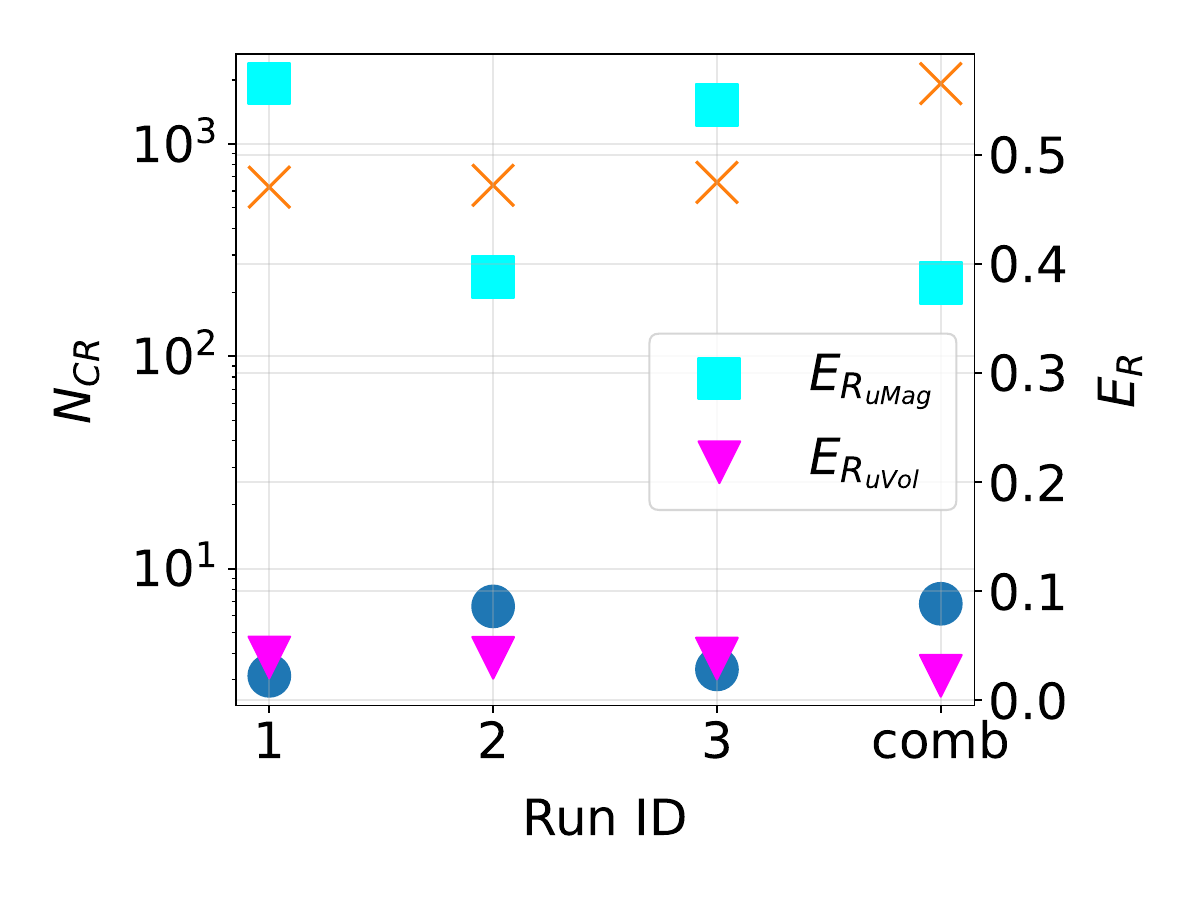}\\
    \includegraphics[width=0.49\linewidth]{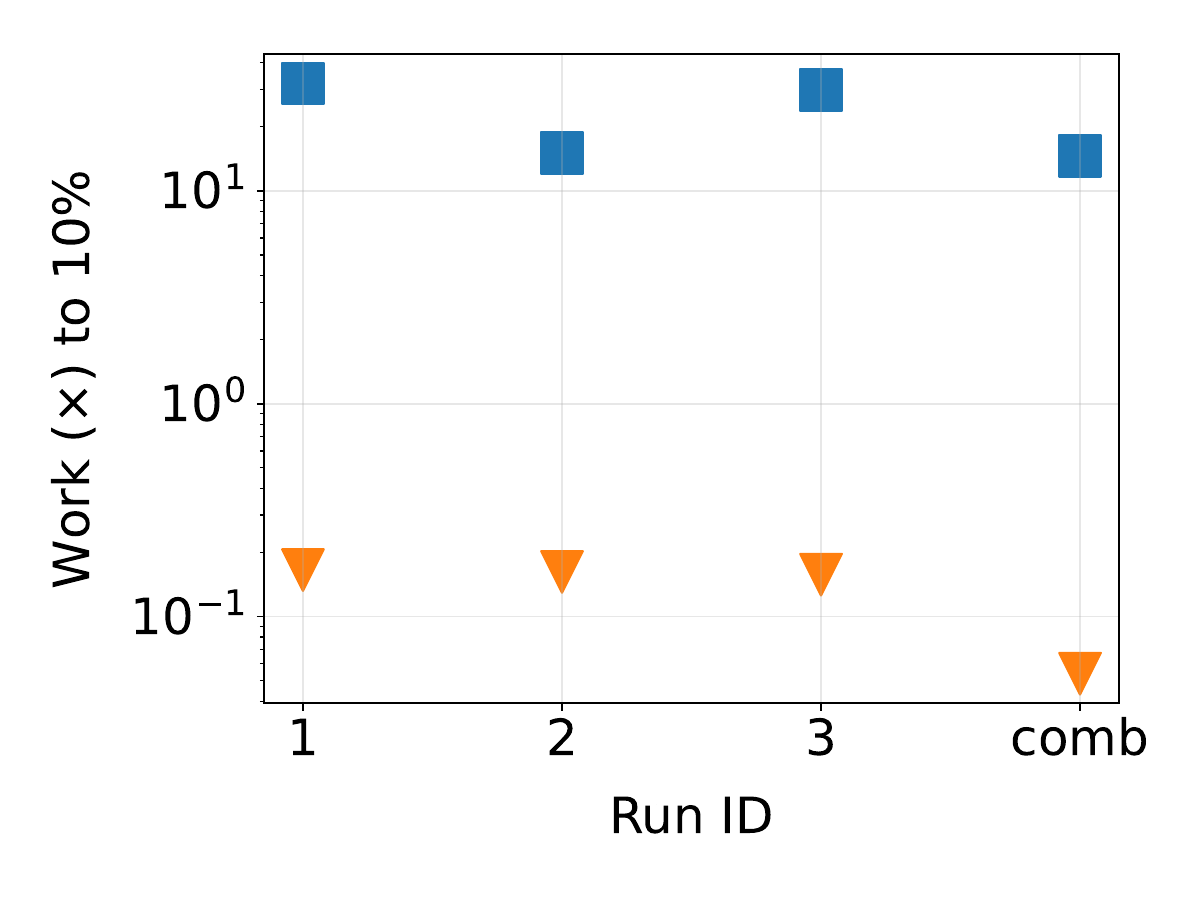}
    \includegraphics[width=0.49\linewidth]{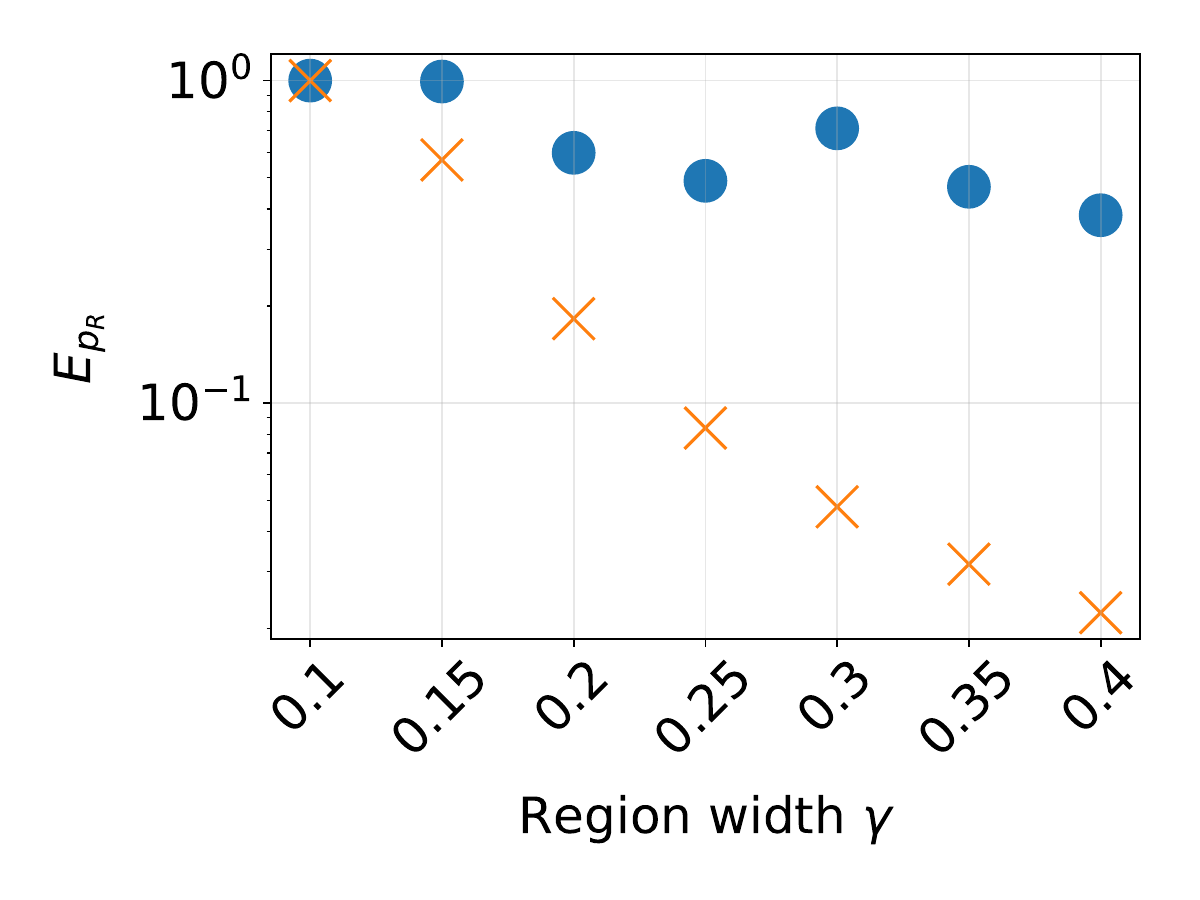}
    
    \caption{Region diagnostics for the uniform-in-magnitude (blue) and uniform-in-volume (orange) priors, across the PE runs for GW250114, which are labelled on the x-axis by the run id. Also shown are the data points for the combined set of samples from across the runs. The regions chosen are $R_{umag} \in \{\theta_i | \chi_{1i},\chi_{2i} > 0.6 \}$ and $R_{uvol} in \{ \theta_i | \chi_{1i},\chi_{2i} < 0.4 \} $, respectively. The top left plot shows the estimated posterior mass in the region $R$. The left axis of the top right plot shows the effective number of equal-weighted samples present in that set that describes the conditional distribution in that region. The right axis shows the relative error in the estimation of $\hat{p}_R$. The bottom left plot shows the amount of work (as a factor) required to attain a target $10\%$ relative error in the estimator $\hat{p}_R$. The bottom right plot shows the relative errors $E_R$ for a range of widths $\gamma \in [0.05, 0.4] $, so that the lower cutoff spin of the chosen region for the uniform-in-magnitude diagnostics is $1-\gamma$ and the upper cutoff for the uniform-in-volume priors is $\gamma$.}
    \label{fig:diag}
\end{figure*}

Apart from the question of motivation for agnosticism in the choice of priors, they can also determine how well the parameter space is sampled. As mentioned earlier, the uniform-in-magnitude resolves the region of the low-spin parameter space better than the uniform-in-volume prior, while underexploring the high-spin states. This becomes evident when one attempts to use re-weighting of the posterior samples from the uniform-in-magnitude run to obtain the posterior samples for the uniform-in-volume priors. Fig.~\ref{fig:reweight} shows the performance of the reweighting procedure applied to the marginal distributions of the spin magnitudes. This clearly shows the missing posterior mass at higher spin states when the uniform-in-magnitude samples are reweighted (blue) to obtain the uniform-in-volume posteriors. In order to analyze the sufficiency of the samples in this region of the parameter space further, we define a mask $I(\theta)$ that is zero outside of the region under consideration and 1 otherwise. Assuming the posterior samples are drawn independently, one can quantify the effective equal-weight number of samples $N_{CR}$ of the conditional posterior distribution $\hat{p}_R (\theta | d), \theta \in R$ that leads to the same variance on a mean estimator as that computed using a set of weighted samples, given the nested samples for $p(\theta)$ \cite{Kish1965SurveySampling}:
\begin{eqnarray}
    N_{CR} = \dfrac{1}{\sum_j W_{(R), j}^2} \label{eqn:ncr}
\end{eqnarray}
Where $W_{(R), j} = I(\theta_j) W(\theta_j)$ are the normalized importance weights of the samples in $R$ i.e. $\sum_j W_{(R), j} =1$, given the weights $w(\theta_j)$ of all samples. 
A low $N_{CR}$ means that a sufficient number of samples are not present to reliably describe the conditional distribution in $R$.

A further independent and useful quantity is the variance of the posterior mass $\hat{p}_R = \sum_j I(\theta_j) p(\theta_j, d)$ in $R$. Again, under the assumption that the samples are identical, independently distributed variables, the posterior mass in $R$ can be treated as the ratio of two random variables $\sum_i w_i I_i$ and $\sum_j w_j$ given the un-normalized weights $w_i$, and can be expanded about their respective means (see end matter for further details), leading to:
\begin{equation}
    var(\hat{p}_R) \approx \sum_i W_i^2 (I(\theta_i) - \hat{p}_R)^2 
\end{equation}
Where $W_i$ are the normalized importance weights of the nested samples.


One can therefore consider two useful diagnostics to assess the reliability of the representation of a region $R$ of the prior space:  (i) $N_{CR}$ that describes the reliability of the \emph{conditional} distribution of $p(\theta | d)$ in $R$. When the posterior mass $\hat{p}_R$ is small, as maybe the case of e.g., $R \equiv \{\theta_i | \chi_{1,2} > 0.6\}$ with uniform-in-magnitude priors (or for small spins with unifom-in-volume priors), it is useful to also compare the standard deviation $std(\hat{p}_R)$ with the value of $\hat{p}_R$ itself, which leads us to the second diagnostic i.e. (ii).  $E_R \equiv \Bar{\delta}\hat{p}_R = std(\hat{p}_R)/\hat{p}_R) \label{eqn:er}$. 

Computing this for \gwthr\, in the region $R_{umag}$ defined by $\chi_{1,2} > 0.6$ for the posterior samples from the uniform-in-magnitude priors, we find that $N_{CR}$ is $O(1)$ (left axis of top right panel of Fig.~\ref{fig:diag}), and $\hat{p}_R$ changes significantly between runs (top left). The relative error estimate (right axis of top right panel) in the posterior mass $\hat{p}_R$ (top left panel) in $R_{umag}$ is also very high, with a value of $E_R \gtrsim O(100)$. Thus, the moderate to high-spin region is incompletely explored with the uniform-in-magnitude priors, and the available number of samples there is too low for reliably estimating the shape of the posterior in $R$, and therefore reweighting cannot be expected to work. One can also ask what amount of work is required, as a factor relative to the current runs with 2000 live points, to achieve a target relative error in $\hat{p}_R$ of $10\%$. This is plotted in the right panel, and shows that more than approximately 10 times the live points are necessary to have a sufficient number of samples in the given region in order to attain this target error. Alternatively, 10 times the runs with the same number of live points could be performed, but the result would be less consistent. This is clearly not feasible and scalable to the sheer number of observed events in O4 Run, let alone in future runs / next generation gravitational detectors. Finally, we report that, using the uniform-in-magnitude priors, restricting the region of interest to $\chi_{1,2}<0.15$ yields similar uncertainties as those of the uniform-in-magnitude priors in the region $\chi_{1,2}>0.6$. In comparison, the uniform-in-volume priors for the region defined by $\chi_{1,2} < \gamma$ have generally stable errors across runs for $\gamma=0.4$, and the region is much better resolved, as suggested by the work required in the bottom left panel being less than unity. A plot of the errors as a function of the region widths, wherein the region corresponds to low spin for uVol and high spin for uMag priors are shown in the bottom right panel. This also suggests that the low-spin region in the uvol priors starts becoming less reliable below spins of $q\approx0.225$, where the errors surpass the $10\%$ threshold. They attain the same errors as those of uMag priors for a lower cutoff of $\chi_{1,2} = 0.6$ when the upper cutoff is $\chi_{1,2} = 0.1$. 

Due to this finite sample-size effect and a lack of sufficient samples in the high-spin region of the parameter space, one cannot obtain the posteriors for uniform-in-volume priors through reweighting, as some regions of parameter space are underexplored. This would be true for any pair of priors, unless they are only marginally different from each other or a sufficiently large number of samples are drawn.
\begin{figure}
    \centering
    \includegraphics[width=\linewidth]{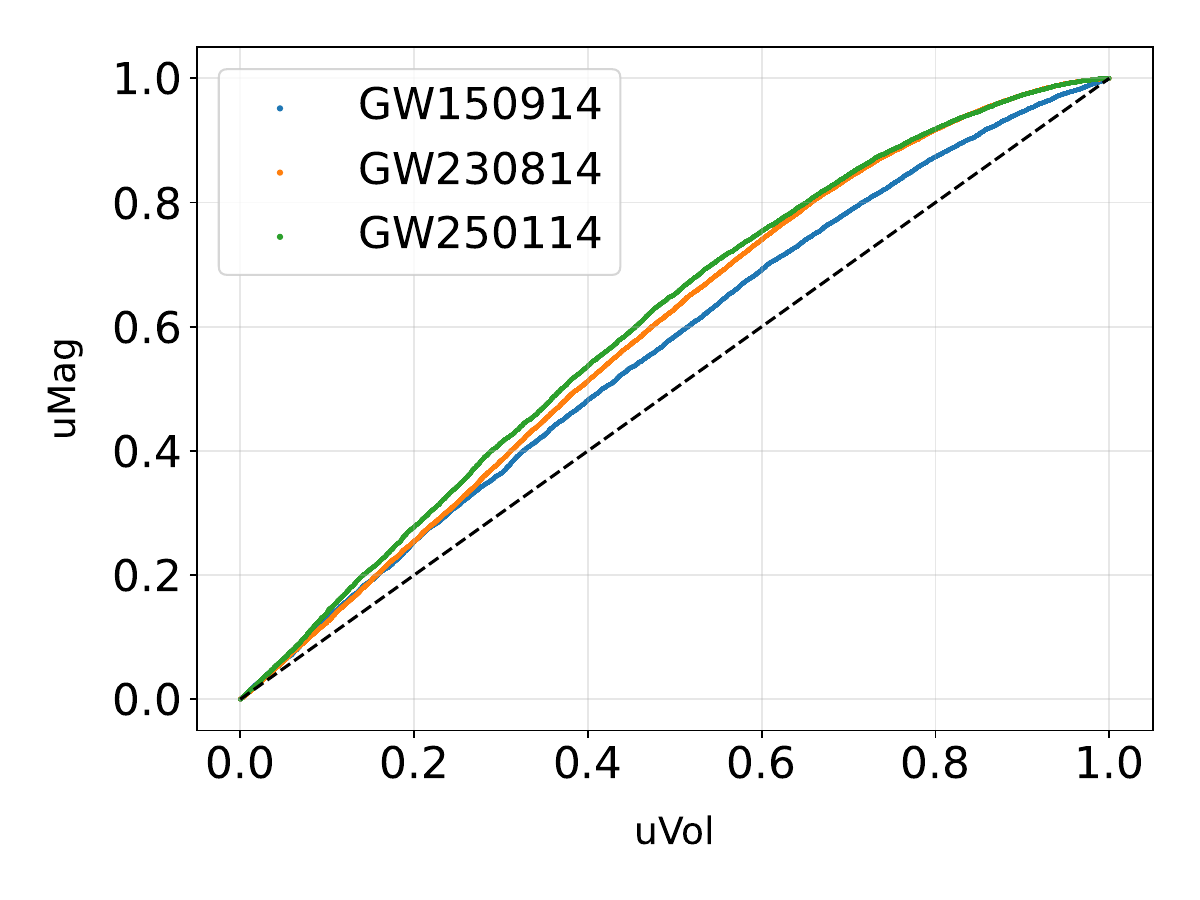}
    \caption{p-p plot for the log-likelihood distributions of the PE runs with uniform-in-magnitude posteriors (y-axis) and uniform-in-volume posteriors (x-axis). This shows that the log-likelihood distributions consistently favour lower values for the uniform-in-magnitude samples.}
    \label{fig:placeholder}
\end{figure}
We find that in all the events analyzed here, the new priors lead to a better accumulation of the posterior mass around the maximum likelihood regions. While the posteriors are always dependent on the priors, in the absence of astrophysical priors, it is important that the chosen prior allows a stochastic sampler to explore the entire parameter space sufficiently, with equal weight to every possible spin state. While the existing priors used in LVK analyses have arguably better sampling resolution at low spins, this letter shows that they are less capable of resolving high-spin regions and can miss out on higher likelihood regions. In contrast, the new priors, we claim, allow the sampler to explore the spin configuration space democratically.

We conclude this letter with the following findings: i). The agnosticity of priors depends on the chosen topology. The LVK spin priors currently being used are only agnostic in a particular topology $S^2 \times I$, and not in the canonical 3D Euclidean space $B^3$. ii). Choosing uniform-in-volume priors consistently returns marginally higher log-likelihood distributions for the events analyzed iii). With the uniform-in-volume priors, stochastic samplers explore the spin-configuration space democratically and are better able to recover the maximum likelihood parameters than the uniform-in-magnitude priors. In particular, in the exceptional events analyzed, they lead to the discovery of samples with a higher likelihood at higher spin states where the uniform-in-magnitude priors have $O(1)$ effective samples and thus are underexplored. iv). Thus, it may be more appropriate to analyze gravitational wave events with alternate power law priors like the uniform-in-volume ($\pi\propto\chi^2$) or $\propto\chi$ prior first, and following it up with the uniform-in-magnitude priors if the posteriors / maxL samples show support at lower spins. This would also ensure that the population inference is not affected by the lack of samples in certain regions of the parameter space.

We reiterate the well-known fact that in Bayesian inference, the posteriors and their sampling processes can be strongly dependent on the prior distribution, especially when the likelihood does not contain sufficient information to transform the priors. In the absence of astrophysical priors for the spins, it is important to choose agnostic priors, so that while we accumulate more events to gain an understanding of the population parameters of the binary compact objects, we do not bias our perception of the universe.

\begin{acknowledgments}
We are grateful to Abhay Ashtekar, Debatri Chattopadhyay, John Veitch, K.G. Arun, Matthew Mold, Thomas Dent, Sharan Banagiri, Thomas Callister, Koustav Chandra, Rossella Gamba, David Radice, and Aditya Vijaykumar for illuminating discussions that helped clarify some of the conceptual points and improve the presentation of this letter. This research is supported by National Science Foundation grants Nos.\, AST-2307147, PHY-2207638, PHY-2308886, and No.~PHY-2309064.

This work was also supported by the National Science Foundation awards PHY-2309064 and PHYS-2308886.
This material is based upon work supported by NSF's LIGO Laboratory, which is a major facility fully funded by the National Science Foundation

Facilities: EGO:Virgo, GEO600, Kamioka:KAGRA, LIGO

Software: Plots were prepared with Matplotlib \cite{Hunter2007Matplotlib}, waveforms were generated through LALSuite \cite{LVK2018LALSuite, Wette2020SWIGLAL}.  NumPy \cite{harris2020array}, SciPy 
\cite{Virtanen2020SciPy} were used for data processing in generating the figures
and quantities in the manuscript. Bilby \cite{Ashton2019Bilby, RomeroShaw2020BilbyValidation, Smith2020pBilby} and Dynesty \cite{Speagle2020dynesty, Koposov2025dynesty_v3_0_0} for stochastic sampling with Bayesian Inference.

Computing: The authors acknowledge the computational resources provided by the LIGO Laboratory’s CIT cluster, which is supported by National Science Foundation Grants PHY-0757058 and PHY-0823459, and the Gwave cluster at Pennsylvania State University (supported by NSF grants OAC-2346596, OAC-2201445, OAC-2103662, OAC-2018299, and PHY-2110594).
\end{acknowledgments}

\bibliography{references}

@String{apj =    "Astrophys.\ J."}

@String{arxiv =        "http://www.arxiv.org/abs/"}

@article{AdvancedLIGOCQG2015,
  author={{The LIGO Scientific Collaboration}},
  title={{Advanced LIGO}},
  journal={Classical and Quantum Gravity},
  volume={32},
  number={7},
  pages={074001},
  url={http://stacks.iop.org/0264-9381/32/i=7/a=074001},
  year={2015}
}

@ARTICLE{AdvancedVirgoCQG2015,
   author = {{Acernese}, F. and {Agathos}, M. and {Agatsuma}, K. and {Aisa}, D. and 
	{Allemandou}, N. and {Allocca}, A. and {Amarni}, J. and {Astone}, P. and 
	{Balestri}, G. and {Ballardin}, G. and et al.},
    title = "{Advanced Virgo: a second-generation interferometric gravitational wave detector}",
  journal = {Classical and Quantum Gravity},
archivePrefix = "arXiv",
   eprint = {1408.3978},
 primaryClass = "gr-qc",
     year = 2015,
    month = jan,
   volume = 32,
   number = 2,
      eid = {024001},
    pages = {024001},
      doi = {10.1088/0264-9381/32/2/024001},
   adsurl = {http://adsabs.harvard.edu/abs/2015CQGra..32b4001A}
}

@ARTICLE{KAGRA2012,
   author = {{Somiya}, K.},
    title = "{Detector configuration of KAGRA-the Japanese cryogenic gravitational-wave detector}",
  journal = {Classical and Quantum Gravity},
archivePrefix = "arXiv",
   eprint = {1111.7185},
 primaryClass = "gr-qc",
     year = 2012,
    month = jun,
   volume = 29,
   number = 12,
      eid = {124007},
    pages = {124007},
      doi = {10.1088/0264-9381/29/12/124007},
   adsurl = {http://adsabs.harvard.edu/abs/2012CQGra..29l4007S}
}

@article{Mandel:2018mve,
  author       = {Mandel, Ilya and Broekgaarden, Floor S.},
  title        = {Merging stellar-mass binary black holes},
  journal      = {Phys. Rept.},
  volume       = {955},
  pages        = {1--63},
  year         = {2022},
  doi          = {10.1016/j.physrep.2022.01.003},
  eprint       = {1806.05820},
  archivePrefix= {arXiv},
  primaryClass = {astro-ph.HE}
}

@article{Mandel:2021smh,
  author       = {Mandel, Ilya and Broekgaarden, Floor S.},
  title        = {Rates of Compact Object Coalescences},
  journal      = {Living Rev. Rel.},
  volume       = {25},
  number       = {1},
  pages        = {1},
  year         = {2022},
  doi          = {10.1007/s41114-022-00040-1},
  eprint       = {2107.14239},
  archivePrefix= {arXiv},
  primaryClass = {astro-ph.HE}
}

@article{Mapelli:2020vca,
  author       = {Mapelli, Michela},
  title        = {Binary Black Hole Mergers: Formation and Populations},
  journal      = {Front. Astron. Space Sci.},
  volume       = {7},
  pages        = {38},
  year         = {2020},
  doi          = {10.3389/fspas.2020.00038}
}

@article{Postnov:2014tza,
  author       = {Postnov, Konstantin A. and Yungelson, Lev R.},
  title        = {The Evolution of Compact Binary Star Systems},
  journal      = {Living Rev. Rel.},
  volume       = {17},
  pages        = {3},
  year         = {2014},
  doi          = {10.12942/lrr-2014-3}
}

@unpublished{LIGOScientific:2025pvj,
    author = "Abac, A. G. and others",
    collaboration = "LIGO Scientific, VIRGO, KAGRA",
    title = "{GWTC-4.0: Population Properties of Merging Compact Binaries}",
    eprint = "2508.18083",
    archivePrefix = "arXiv",
    primaryClass = "astro-ph.HE",
    reportNumber = "LIGO-P2400004",
    month = "8",
    note = "Submitted for publication to ApJ Lett.",
    year = "2025"
}

@unpublished{LIGOScientific:2025hdt,
    author = "Abac, A. G. and others",
    collaboration = "LIGO Scientific, VIRGO, KAGRA",
    title = "{GWTC-4.0: An Introduction to Version 4.0 of the Gravitational-Wave Transient Catalog}",
    eprint = "2508.18080",
    archivePrefix = "arXiv",
    primaryClass = "gr-qc",
    reportNumber = "LIGO-P2400293",
    month = "8",
    note = "Submitted for publication to ApJ Lett.",
    year = "2025"
}

@unpublished{LIGOScientific:2025slb,
    author = "Abac, A. G. and others",
    collaboration = "LIGO Scientific, VIRGO, KAGRA",
    title = "{GWTC-4.0: Updating the Gravitational-Wave Transient Catalog with Observations from the First Part of the Fourth LIGO-Virgo-KAGRA Observing Run}",
    eprint = "2508.18082",
    archivePrefix = "arXiv",
    primaryClass = "gr-qc",
    reportNumber = "LIGO-P2400386",
    month = "8",
    note = "Submitted for publication to ApJ Lett.",
    year = "2025"
}

@article{Blanchet:2013haa,
    author = "Blanchet, Luc",
    title = "{Post-Newtonian Theory for Gravitational Waves}",
    eprint = "1310.1528",
    archivePrefix = "arXiv",
    primaryClass = "gr-qc",
    doi = "10.12942/lrr-2014-2",
    journal = "Living Rev. Rel.",
    volume = "17",
    pages = "2",
    year = "2014"
}

@article{Baird:2012cu,
    author = "Baird, Emily and Fairhurst, Stephen and Hannam, Mark and Murphy, Patricia",
    title = "{Degeneracy between mass and spin in black-hole-binary waveforms}",
    eprint = "1211.0546",
    archivePrefix = "arXiv",
    primaryClass = "gr-qc",
    doi = "10.1103/PhysRevD.87.024035",
    journal = "Phys. Rev. D",
    volume = "87",
    number = "2",
    pages = "024035",
    year = "2013"
}

@article{Talbot:2017yur,
  author         = "Talbot, Colm and Thrane, Eric",
  title          = "{Determining the population properties of spinning black holes}",
  journal        = "Phys. Rev. D",
  volume         = "96",
  number         = "2",
  pages          = "023012",
  year           = "2017",
  doi            = "10.1103/PhysRevD.96.023012",
  eprint         = "1704.08370",
  archivePrefix  = "arXiv",
  primaryClass   = "astro-ph.HE"
}

@article{Gerosa:2017kvu,
  author         = "Gerosa, Davide and Berti, Emanuele",
  title          = "{Are merging black holes born from stellar collapse or previous mergers?}",
  journal        = "Phys. Rev. D",
  volume         = "95",
  number         = "12",
  pages          = "124046",
  year           = "2017",
  doi            = "10.1103/PhysRevD.95.124046",
  eprint         = "1703.06223",
  archivePrefix  = "arXiv",
  primaryClass   = "gr-qc"
}

@article{Kidder:1995zr,
  author         = "Kidder, Lawrence E.",
  title          = "{Spin effects in the phasing of gravitational waves from inspiraling compact binaries}",
  journal        = "Phys. Rev. D",
  volume         = "52",
  pages          = "821--847",
  year           = "1995",
  doi            = "10.1103/PhysRevD.52.821",
  eprint         = "gr-qc/9506022",
  archivePrefix  = "arXiv",
  primaryClass   = "gr-qc"
}

@article{Berti:2009kk,
  author         = "Berti, Emanuele and Cardoso, Vitor and Starinets, Andrei O.",
  title          = "{Quasinormal modes of black holes and black branes}",
  journal        = "Class. Quant. Grav.",
  volume         = "26",
  pages          = "163001",
  year           = "2009",
  doi            = "10.1088/0264-9381/26/16/163001",
  eprint         = "0905.2975",
  archivePrefix  = "arXiv",
  primaryClass   = "hep-th"
}

@article{Moore:2021ibq,
  author         = "Moore, Christopher J. and Finch, Eliot and Buscicchio, Riccardo and Gerosa, Davide",
  title          = "{Testing general relativity with gravitational-wave catalogs: The insidious nature of waveform systematics}",
  journal        = "iScience",
  volume         = "24",
  number         = "7",
  pages          = "102577",
  year           = "2021",
  doi            = "10.1016/j.isci.2021.102577",
  eprint         = "2103.16486",
  archivePrefix  = "arXiv",
  primaryClass   = "gr-qc"
}

@article{Thorne:1974ve,
  author         = "Thorne, Kip S.",
  title          = "{Disk accretion onto a black hole. II. Evolution of the hole}",
  journal        = "Astrophys. J.",
  volume         = "191",
  pages          = "507--520",
  year           = "1974",
  doi            = "10.1086/152991"
}

@article{Kesden:2008ga,
  author         = "Kesden, Michael",
  title          = "{Can binary mergers produce maximally spinning black holes?}",
  journal        = "Phys. Rev. D",
  volume         = "78",
  pages          = "084030",
  year           = "2008",
  doi            = "10.1103/PhysRevD.78.084030",
  eprint         = "0807.3043",
  archivePrefix  = "arXiv",
  primaryClass   = "astro-ph"
}

@article{Kimball:2020qyd,
  author         = "Kimball, Christopher and others",
  title          = "{Evidence for hierarchical black hole mergers in the second LIGO–Virgo gravitational-wave catalog}",
  journal        = "Astrophys. J. Lett.",
  volume         = "915",
  number         = "2",
  pages          = "L35",
  year           = "2021",
  doi            = "10.3847/2041-8213/ac0aef",
  eprint         = "2011.05332",
  archivePrefix  = "arXiv",
  primaryClass   = "astro-ph.HE"
}

@article{Tagawa:2021ofj,
  author         = "Tagawa, Hiromichi and Kocsis, Bence and Saitoh, Takayuki R. and others",
  title          = "{Signatures of hierarchical mergers in black hole spin and mass distributions}",
  journal        = "Mon. Not. Roy. Astron. Soc.",
  volume         = "508",
  number         = "3",
  pages          = "L41--L46",
  year           = "2021",
  doi            = "10.1093/mnrasl/slab108",
  eprint         = "2104.09510",
  archivePrefix  = "arXiv",
  primaryClass   = "astro-ph.HE"
}

@article{Mahapatra:2022ngs,
    author = "Mahapatra, Parthapratim and Chattopadhyay, Debatri and Gupta, Anuradha and Favata, Marc and Sathyaprakash, B. S. and Arun, K. G.",
    title = "{Predictions of a simple parametric model of hierarchical black hole mergers}",
    eprint = "2209.05766",
    archivePrefix = "arXiv",
    primaryClass = "astro-ph.HE",
    reportNumber = "LIGO preprint number P2200265",
    doi = "10.1103/PhysRevD.111.023013",
    journal = "Phys. Rev. D",
    volume = "111",
    number = "2",
    pages = "023013",
    year = "2025"
}

@unpublished{Gupta:2024gun,
    author = "Gupta, Anuradha and others",
    title = "{Possible causes of false general relativity violations in gravitational wave observations}",
    eprint = "2405.02197",
    archivePrefix = "arXiv",
    primaryClass = "gr-qc",
    note = {10.21468/SciPostPhysCommRep.5},
    month = "5",
    year = "2024"
}

@misc{lange2018rapidaccurateparameterinference,
      title={Rapid and accurate parameter inference for coalescing, precessing compact binaries}, 
      author={Jacob Lange and Richard O'Shaughnessy and Monica Rizzo},
      year={2018},
      eprint={1805.10457},
      archivePrefix={arXiv},
      primaryClass={gr-qc},
      url={https://arxiv.org/abs/1805.10457}, 
}

@article{GWTC1,
  title = {GWTC-1: A Gravitational-Wave Transient Catalog of Compact Binary Mergers Observed by LIGO and Virgo during the First and Second Observing Runs},
  author = {Abbott, B. P. et. al},
  collaboration = {LIGO Scientific Collaboration and Virgo Collaboration},
  journal = {Phys. Rev. X},
  volume = {9},
  issue = {3},
  pages = {031040},
  numpages = {49},
  year = {2019},
  month = {Sep},
  publisher = {American Physical Society},
  doi = {10.1103/PhysRevX.9.031040},
  url = {https://link.aps.org/doi/10.1103/PhysRevX.9.031040}
}

@article{Chatziioannou:2018wqx,
    author = "Chatziioannou, Katerina and Lovelace, Geoffrey and Boyle, Michael and Giesler, Matthew and Hemberger, Daniel A. and Katebi, Reza and Kidder, Lawrence E. and Pfeiffer, Harald P. and Scheel, Mark A. and Szil{\'a}gyi, B{\'e}la",
    title = "{Measuring the properties of nearly extremal black holes with gravitational waves}",
    eprint = "1804.03704",
    archivePrefix = "arXiv",
    primaryClass = "gr-qc",
    doi = "10.1103/PhysRevD.98.044028",
    journal = "Phys. Rev. D",
    volume = "98",
    number = "4",
    pages = "044028",
    year = "2018"
}

@article{DeRenzis:2022vsj,
    author = "De Renzis, Viola and Gerosa, Davide and Pratten, Geraint and Schmidt, Patricia and Mould, Matthew",
    title = "{Characterization of merging black holes with two precessing spins}",
    eprint = "2207.00030",
    archivePrefix = "arXiv",
    primaryClass = "gr-qc",
    doi = "10.1103/PhysRevD.106.084040",
    journal = "Phys. Rev. D",
    volume = "106",
    number = "8",
    pages = "084040",
    year = "2022"
}

@article{Vitale:2017cfs,
    author = "Vitale, Salvatore and Gerosa, Davide and Haster, Carl-Johan and Chatziioannou, Katerina and Zimmerman, Aaron",
    title = "{Impact of Bayesian Priors on the Characterization of Binary Black Hole Coalescences}",
    eprint = "1707.04637",
    archivePrefix = "arXiv",
    primaryClass = "gr-qc",
    reportNumber = "LIGO-P1700176, LIGO P1700176",
    doi = "10.1103/PhysRevLett.119.251103",
    journal = "Phys. Rev. Lett.",
    volume = "119",
    number = "25",
    pages = "251103",
    year = "2017"
}

@article{Callister:2022qwb,
    author = "Callister, Thomas A. and Miller, Simona J. and Chatziioannou, Katerina and Farr, Will M.",
    title = "{No Evidence that the Majority of Black Holes in Binaries Have Zero Spin}",
    eprint = "2205.08574",
    archivePrefix = "arXiv",
    primaryClass = "astro-ph.HE",
    doi = "10.3847/2041-8213/ac847e",
    journal = "Astrophys. J. Lett.",
    volume = "937",
    number = "1",
    pages = "L13",
    year = "2022"
}

@article{Tong:2022iws,
    author = "Tong, Hui and Galaudage, Shanika and Thrane, Eric",
    title = "{Population properties of spinning black holes using the gravitational-wave transient catalog 3}",
    eprint = "2209.02206",
    archivePrefix = "arXiv",
    primaryClass = "astro-ph.HE",
    doi = "10.1103/PhysRevD.106.103019",
    journal = "Phys. Rev. D",
    volume = "106",
    number = "10",
    pages = "103019",
    year = "2022"
}

@unpublished{LIGOScientific:2025yae,
    author = "Abac, A. G. and others",
    collaboration = "LIGO Scientific, VIRGO, KAGRA",
    title = "{GWTC-4.0: Methods for Identifying and Characterizing Gravitational-wave Transients}",
    eprint = "2508.18081",
    archivePrefix = "arXiv",
    primaryClass = "gr-qc",
    reportNumber = "LIGO-P2400300",
    month = "8",
    year = "2025",
    note={2508.18081}
}

@article{Banagiri:2025dxo,
    author = "Banagiri, Sharan and Callister, Thomas A. and Adamcewicz, Christian and Doctor, Zoheyr and Kalogera, Vicky",
    title = "{Structure and Skewness of the Effective Inspiral Spin Distribution of Binary Black Hole Mergers}",
    eprint = "2501.06712",
    archivePrefix = "arXiv",
    primaryClass = "astro-ph.HE",
    doi = "10.3847/1538-4357/adf4c6",
    journal = "Astrophys. J.",
    volume = "990",
    number = "2",
    pages = "147",
    year = "2025"
}

@article{Adamcewicz:2025phm,
    author = "Adamcewicz, Christian and Guttman, Nir and Lasky, Paul D. and Thrane, Eric",
    title = "{Do Both Black Holes Spin in Merging Binaries? Evidence from GWTC-4 and Astrophysical Implications}",
    eprint = "2509.04706",
    archivePrefix = "arXiv",
    primaryClass = "astro-ph.HE",
    doi = "10.3847/1538-4357/ae1370",
    journal = "Astrophys. J.",
    volume = "994",
    number = "2",
    pages = "261",
    year = "2025"
}

@article{Hunter2007Matplotlib,
  author  = {Hunter, John D.},
  title   = {Matplotlib: A 2D Graphics Environment},
  journal = {Computing in Science \& Engineering},
  year    = {2007},
  volume  = {9},
  number  = {3},
  pages   = {90--95},
  doi     = {10.1109/MCSE.2007.55},
}

@misc{LVK2018LALSuite,
  author       = {{LIGO Scientific Collaboration} and {Virgo Collaboration} and {KAGRA Collaboration}},
  title        = {{LVK Algorithm Library - LALSuite}},
  year         = {2018},
  doi          = {10.7935/GT1W-FZ16},
  url          = {https://doi.org/10.7935/GT1W-FZ16},
  note         = {Free software (GPL)}
}

@article{Wette2020SWIGLAL,
  author       = {Wette, Karl},
  title        = {SWIGLAL: Python and Octave interfaces to the LALSuite gravitational-wave data analysis libraries},
  journal      = {SoftwareX},
  volume       = {12},
  year         = {2020},
  pages        = {100634},
  doi          = {10.1016/j.softx.2020.100634},
  url          = {https://doi.org/10.1016/j.softx.2020.100634}
}

@Article{harris2020array,
 title         = {Array programming with {NumPy}},
 author        = {Charles R. Harris and K. Jarrod Millman and St{\'{e}}fan J.
                 van der Walt and Ralf Gommers and Pauli Virtanen and David
                 Cournapeau and Eric Wieser and Julian Taylor and Sebastian
                 Berg and Nathaniel J. Smith and Robert Kern and Matti Picus
                 and Stephan Hoyer and Marten H. van Kerkwijk and Matthew
                 Brett and Allan Haldane and Jaime Fern{\'{a}}ndez del
                 R{\'{i}}o and Mark Wiebe and Pearu Peterson and Pierre
                 G{\'{e}}rard-Marchant and Kevin Sheppard and Tyler Reddy and
                 Warren Weckesser and Hameer Abbasi and Christoph Gohlke and
                 Travis E. Oliphant},
 year          = {2020},
 month         = sep,
 journal       = {Nature},
 volume        = {585},
 number        = {7825},
 pages         = {357--362},
 doi           = {10.1038/s41586-020-2649-2},
 publisher     = {Springer Science and Business Media {LLC}},
 url           = {https://doi.org/10.1038/s41586-020-2649-2}
}

@article{Ashton2019Bilby,
  author       = {Ashton, Gregory and others},
  title        = {{BILBY}: A user-friendly Bayesian inference library for gravitational-wave astronomy},
  journal      = {The Astrophysical Journal Supplement Series},
  year         = {2019},
  volume       = {241},
  number       = {2},
  pages        = {27},
  doi          = {10.3847/1538-4365/ab06fc},
  eprint       = {1811.02042},
  archivePrefix= {arXiv},
  primaryClass = {astro-ph.IM},
  url          = {https://doi.org/10.3847/1538-4365/ab06fc}
}

@article{RomeroShaw2020BilbyValidation,
  author       = {Romero-Shaw, I. M. and Talbot, C. and Biscoveanu, S. and others},
  title        = {Bayesian inference for compact binary coalescences with bilby: validation and application to the first {LIGO}--{Virgo} gravitational-wave transient catalogue},
  journal      = {Monthly Notices of the Royal Astronomical Society},
  year         = {2020},
  volume       = {499},
  number       = {3},
  pages        = {3295--3319},
  doi          = {10.1093/mnras/staa2850},
  url          = {https://doi.org/10.1093/mnras/staa2850}
}

@article{Smith2020pBilby,
  author       = {Smith, Rory J. E. and Ashton, Gregory and Vajpeyi, Avi and Talbot, Colm},
  title        = {Massively parallel Bayesian inference for transient gravitational-wave astronomy},
  journal      = {Monthly Notices of the Royal Astronomical Society},
  year         = {2020},
  volume       = {498},
  number       = {3},
  pages        = {4492--4502},
  doi          = {10.1093/mnras/staa2483},
  url          = {https://doi.org/10.1093/mnras/staa2483}
}

@article{Virtanen2020SciPy,
  author  = {Virtanen, Pauli and Gommers, Ralf and Oliphant, Travis E. and Haberland, Matt and Reddy, Tyler and Cournapeau, David and Burovski, Evgeni and Peterson, Pearu and Weckesser, Warren and Bright, Jonathan and {van der Walt}, St{\'e}fan J. and Brett, Matthew and Wilson, Joshua and Millman, K. Jarrod and Mayorov, Nikolay and Nelson, Andrew R. J. and Jones, Eric and Kern, Robert and Larson, Eric and Carey, C. J. and Polat, {\.I}lhan and Feng, Yu and Moore, Eric W. and VanderPlas, Jake and Laxalde, Denis and Perktold, Josef and Cimrman, Robert and Henriksen, Ian and Quintero, E. A. and Harris, Charles R. and Archibald, Anne M. and Ribeiro, Ant{\^o}nio H. and Pedregosa, Fabian and {van Mulbregt}, Paul and {SciPy 1.0 Contributors}},
  title   = {{SciPy} 1.0: Fundamental Algorithms for Scientific Computing in Python},
  journal = {Nature Methods},
  year    = {2020},
  volume  = {17},
  number  = {3},
  pages   = {261--272},
  doi     = {10.1038/s41592-019-0686-2},
  url     = {https://doi.org/10.1038/s41592-019-0686-2}
}

@article{Speagle2020dynesty,
  author       = {Speagle, Joshua S.},
  title        = {dynesty: a dynamic nested sampling package for estimating Bayesian posteriors and evidences},
  journal      = {Monthly Notices of the Royal Astronomical Society},
  year         = {2020},
  volume       = {493},
  number       = {3},
  pages        = {3132--3158},
  doi          = {10.1093/mnras/staa278},
  eprint       = {1904.02180},
  archivePrefix= {arXiv},
  primaryClass = {astro-ph.IM},
  url          = {https://doi.org/10.1093/mnras/staa278}
}

@misc{Koposov2025dynesty_v3_0_0,
  author       = {Koposov, Sergey and Speagle, Josh and Barbary, Kyle and others},
  title        = {{joshspeagle/dynesty: v3.0.0}},
  year         = {2025},
  month        = oct,
  version      = {v3.0.0},
  publisher    = {Zenodo},
  doi          = {10.5281/zenodo.17268284},
  url          = {https://doi.org/10.5281/zenodo.17268284}
}

@book{Kish1965SurveySampling,
  author    = {Kish, Leslie},
  title     = {Survey Sampling},
  publisher = {John Wiley \& Sons},
  year      = {1965},
  address   = {New York},
}
\section{End Matter}

%
\begin{figure*}
    \centering
    \includegraphics[width=0.49\linewidth]{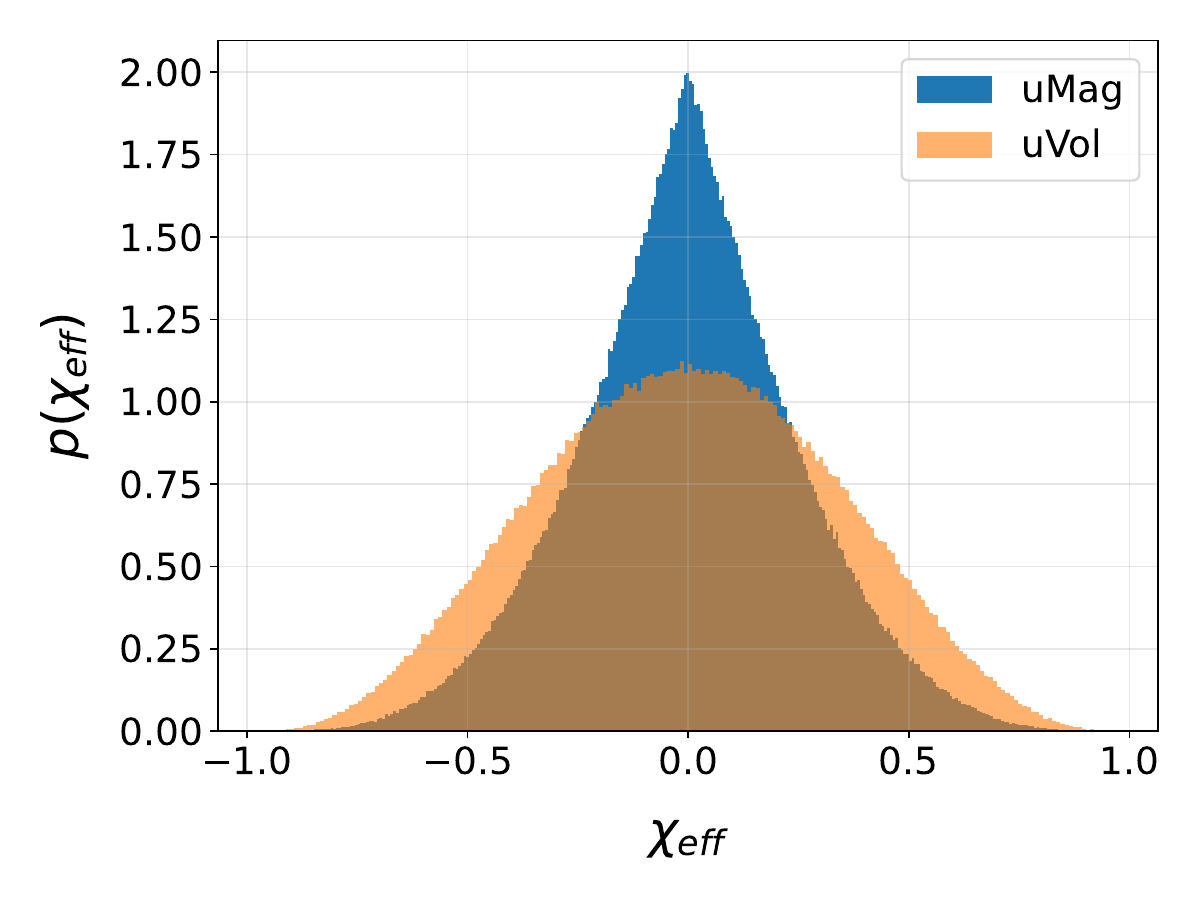}
    \includegraphics[width=0.49\linewidth]{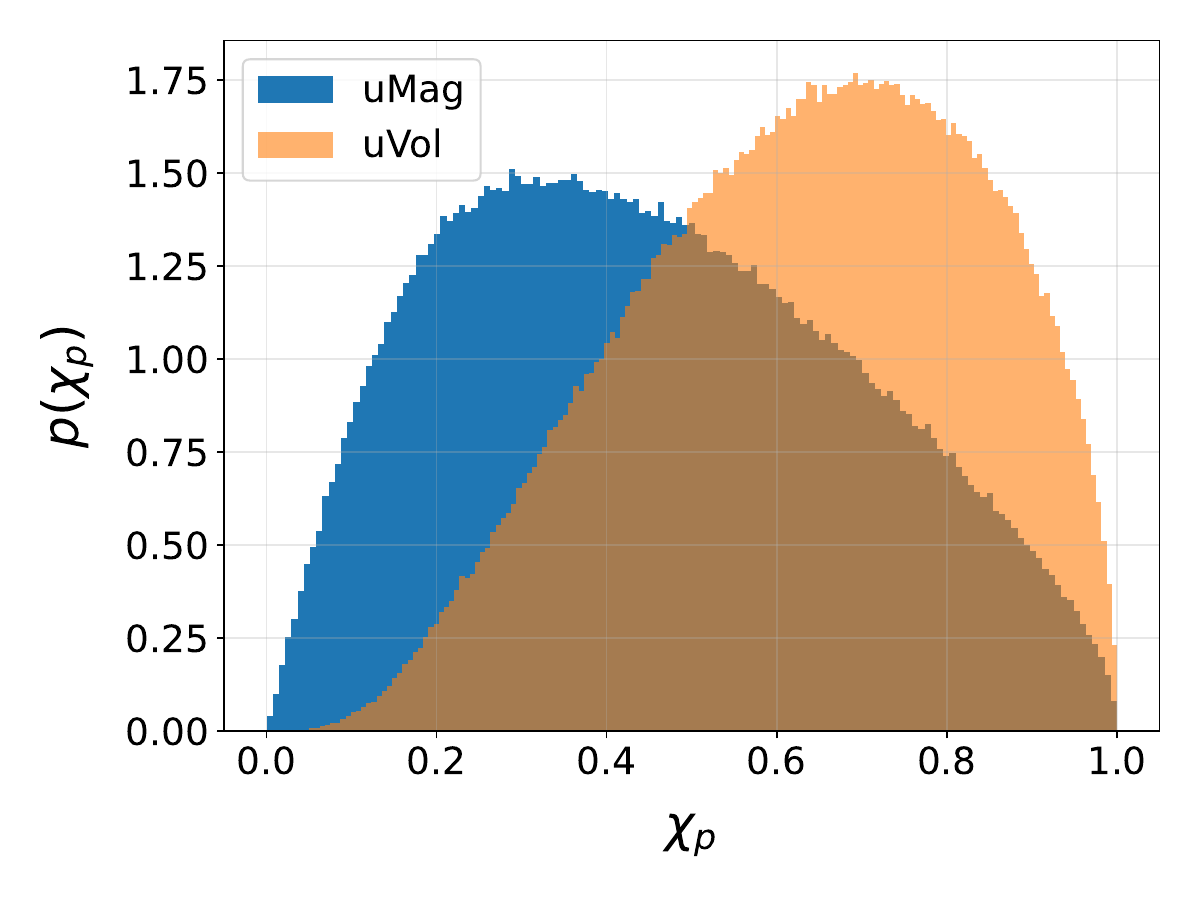}
    \caption{The prior distributions for the effective spin parameter $\chi_{eff}$ (left) and the spin-precession parameter $\chi_p$ (right). }
    \label{fig:eff_priors}
\end{figure*}

\paragraph{Distribution of Cartesian spin components}—For a uniform-in-magnitude prior, the marginal distribution of any Cartesian component exhibits a cusp at zero.  To see this, write the spin vector in spherical coordinates $(\chi, u=\cos\theta, \phi)$, sampled uniformly with $\chi\in[0,1)$, $u\in[-1,1]$, and $\phi\in[0,2\pi)$, and transform to Cartesian components via
\begin{equation}
\chi_x=\chi\sqrt{1-u^2}\cos\phi,\,\,
\chi_y=\chi\sqrt{1-u^2}\sin\phi,\,\,
\chi_z=\chi u.
\label{eq:transformation}
\end{equation}
The Jacobian of this transformation is 
$$\left |\frac{\partial(\chi_x,\chi_y,\chi_z)}{\partial(\chi,u,\phi)} \right |=\chi^2.$$ 
Thus, a flat joint density in $(\chi,u,\phi)$ implies
\begin{equation}
p_M(\boldsymbol{\chi}) \equiv p_M(\chi_x,\chi_y,\chi_z)=\frac{1}{4\pi\,\chi^2}\qquad (|\boldsymbol{\chi}|<1),
\label{eq:pdfMjoint}
\end{equation}
where subscript $M$ corresponds to a distribution in which the spin \textit{magnitude} is chosen to obey a uniform distribution. Marginalizing over $\chi_x$ and $\chi_y$ gives
\begin{equation}
p_M(\chi_z)
= \iint_{\chi_x^2+\chi_y^2 \le 1-\chi_z^2}
\frac{d\chi_x\,d\chi_y}{4\pi\,(\chi_x^2+\chi_y^2+\chi_z^2)},
\quad |\chi_z|<1,
\end{equation}
and $p(\chi_z)=0$ otherwise. Switching to polar coordinates in the $(\chi_x,\chi_y)$-plane with $\rho^2=\chi_x^2+\chi_y^2$,
\begin{eqnarray}
p_M(\chi_z)
& = &  \frac{1}{4\pi}\int_0^{2\pi}\!\!d\phi\int_0^{\sqrt{1-\chi_z^2}} \frac{\rho}{\rho^2+\chi_z^2}\,d\rho \nonumber \\
& =&  -\tfrac{1}{2}\ln|\chi_z|\,, \quad |\chi_z|<1.
\end{eqnarray}
Thus, the marginalized distribution is
\begin{equation}
p_M(\chi_z)= -\tfrac{1}{2}\ln|\chi_z|, \qquad |\chi_z|<1.
\label{eq:pdfMcomponents}
\end{equation}
The density $p_M(\chi_z)$ is an even function of $\chi_z$, vanishes as $|\chi_z|\to 1$, and exhibits a logarithmic (but integrable) cusp at $\chi_z=0$. By rotational symmetry, there is nothing special about the $z$-component: the same one-dimensional marginal applies to any Cartesian component—and, more generally, to the projection of the spin along any fixed direction.

If spin vectors are drawn uniformly within the unit ball, the joint density in Cartesian components is constant over $|\boldsymbol{\chi}|\le 1$ and vanishes outside:
\begin{equation}
p_V(\boldsymbol{\chi})=\frac{3}{4\pi}\,\mathbf{1}, \qquad |\boldsymbol{\chi}|\le 1.
\label{eq:pdfVjoint}
\end{equation}
Subscript $V$ represents distributions where spin \emph{vector} obeys a uniform in volume prior. The one-dimensional marginal for a Cartesian component, say $\chi_z$, follows by integrating over the disk $\rho^2=\chi_x^2+\chi_y^2\le 1-\chi_z^2.$ As before,
\begin{equation}
p_V(\chi_z)=\frac{3}{4\pi}\iint_{\rho^2\le 1-\chi_z^2}\,d\chi_x\,d\chi_y
=\frac{3}{4}\,(1-\chi_z^2).
\label{eq:pdfVcomponents}
\end{equation}
By rotational symmetry, this marginal applies to any Cartesian component. As a consistency check, 
$$\frac{3}{4}\int_{-1}^{1} (1-z^2)\,dz=1.$$ 
The radial distribution is 
\begin{equation}
    p_V(\chi)=3\chi^2, \qquad \chi\in[0,1].
    \label{eq:pdfVmagnitude}
\end{equation}

\paragraph{Sampler and Parameter Estimation setup}- All PE runs were carried out using the Bilby and BilbyPipe infrastructure with the dynesty sampler, 2000 live points, and a stopping criterion of dlogz = 0.01. The official released PSDs and calibration files were used for this purpose.


\begin{table}
\centering
\caption{Summary statistics for 153 binary black holes in GWTC-4 with $\mathrm{FAR}<1\,\mbox{yr}^{-1}.$ Widths are fractional for masses $\Delta X = (X_{\rm upper}-X_{\rm lower})/X$ and absolute for $\chi_{\rm eff}$ $\Delta \chi_{\rm eff} = (\chi_{\rm eff}^{\rm upper}-\chi_{\rm eff}^{\rm lower}).$}
\label{tab:gwtc4-summary}
\begin{tabular}{lcccccc}
\hline
\hline \\[-5pt]
$X$ & $\mbox{Med}(X)$ & $X_{\rm min}$ & $X_{\rm max}$ & $\mbox{Med}({\Delta X})$ & $\Delta X_{\rm min}$ & $\Delta X_{\rm max}$ \\ \\[-5pt]
\hline \\[-7pt]
$m_1$                 & 36   & 8      & 137  & 0.50  & 0.12  & 1.5  \\
$m_2$                 & 26   & 3      & 103  & 0.56  & 0.077 & 1.7  \\
$\mathcal{M}_c$       & 27   & 6      & 102  & 0.28  & 0.018 & 0.86 \\
$M_{\rm tot}$         & 62   & 14     & 238  & 0.32  & 0.061 & 0.87 \\ \\[-7pt]
\hline\\[-5pt]
$\chi_{\rm eff}$      & 0.05 & $-0.3$ & 0.49 & 0.44  & 0.11 & 0.95 \\ \\[-5pt]
\hline
\hline
\end{tabular}
\end{table}

\paragraph{Priors on effective spin-parameters}

The prior distributions for the two spin parameters, the effective spin alignment parameter $\chi_{eff}$ and the spin-precession parameter $\chi_p$ are plotted in Fig.~\ref{fig:eff_priors}. For $\chi_{eff}$, (left panel), it can be seen that the uMag priors are narrower and peaked at 0, with a cusp. In contrast, the uVol priors lead to a less biased prior distribution, with more support at higher values of the magnitude of $\chi_{eff}$, and smooth at 0. For $\chi_p$, the uVol priors are peaked at a larger value of $\chi_p$.

\paragraph{PE differences}
\begin{figure*}
    \centering
    \includegraphics[width=0.32\linewidth]{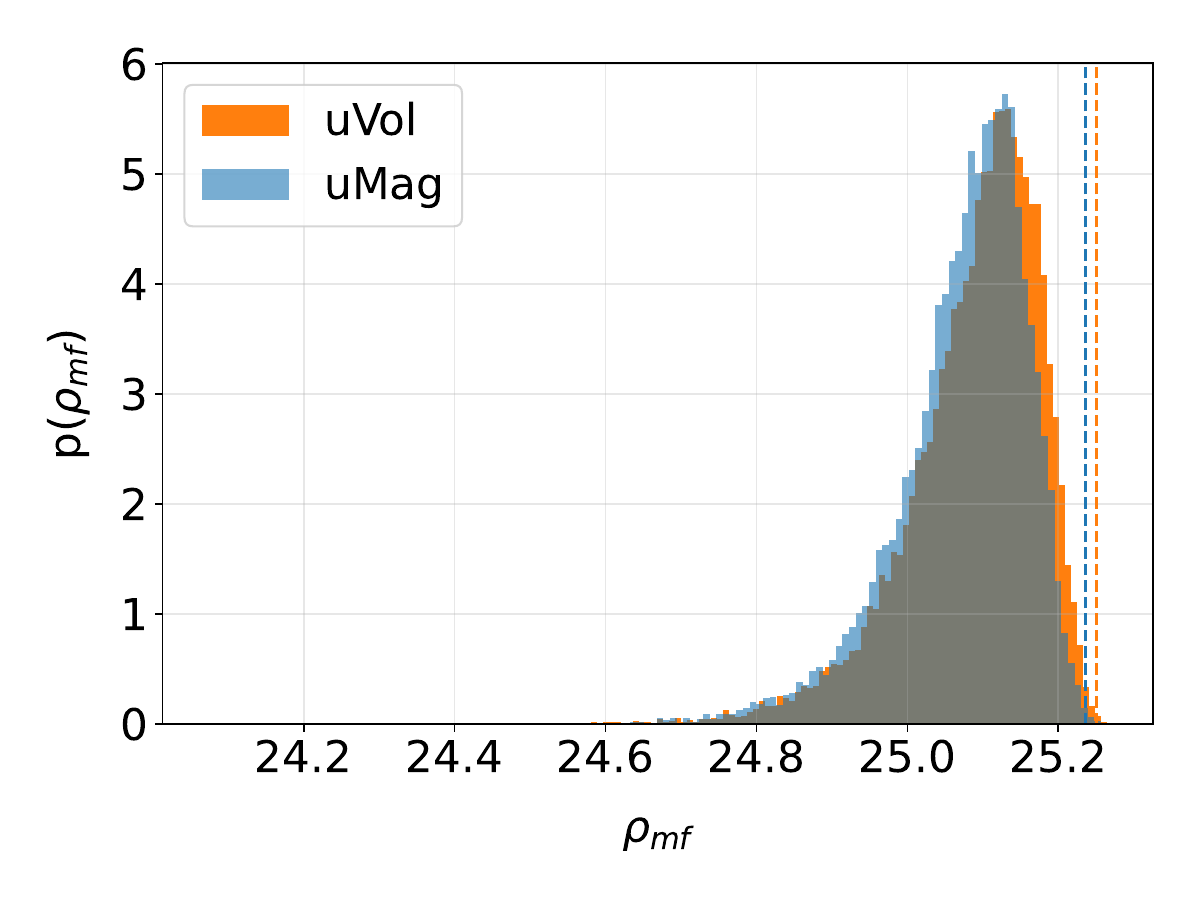}
    \includegraphics[width=0.32\linewidth]{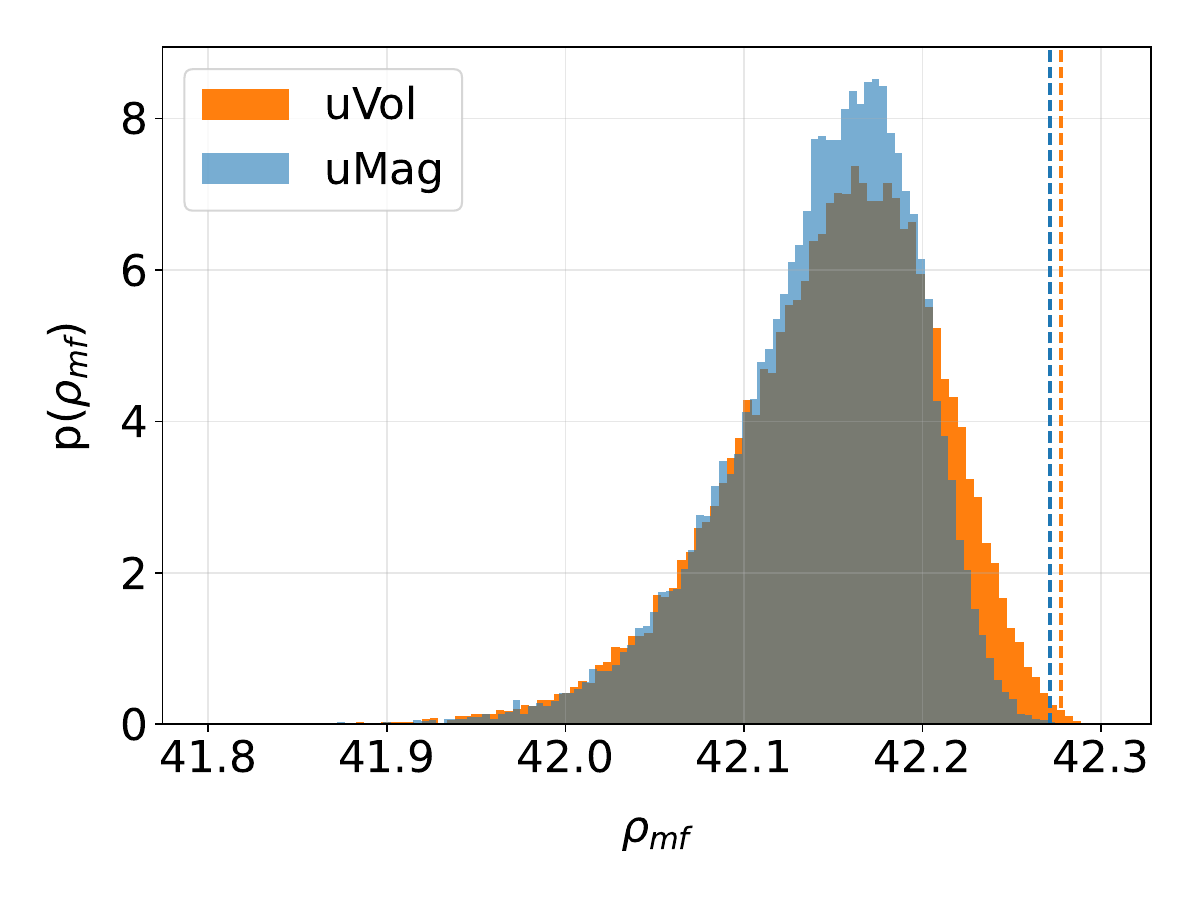}
    \includegraphics[width=0.32\linewidth]{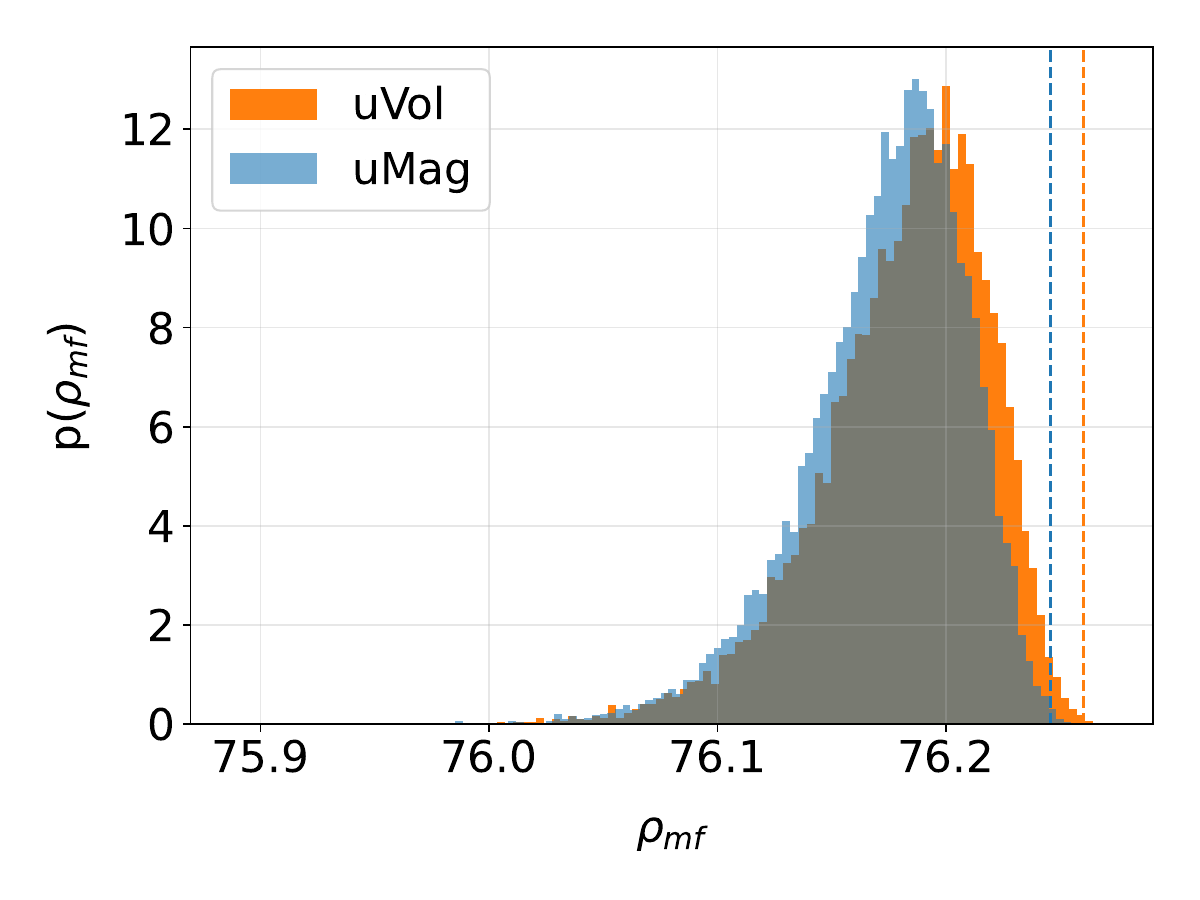}\\
     \includegraphics[width=0.32\linewidth]{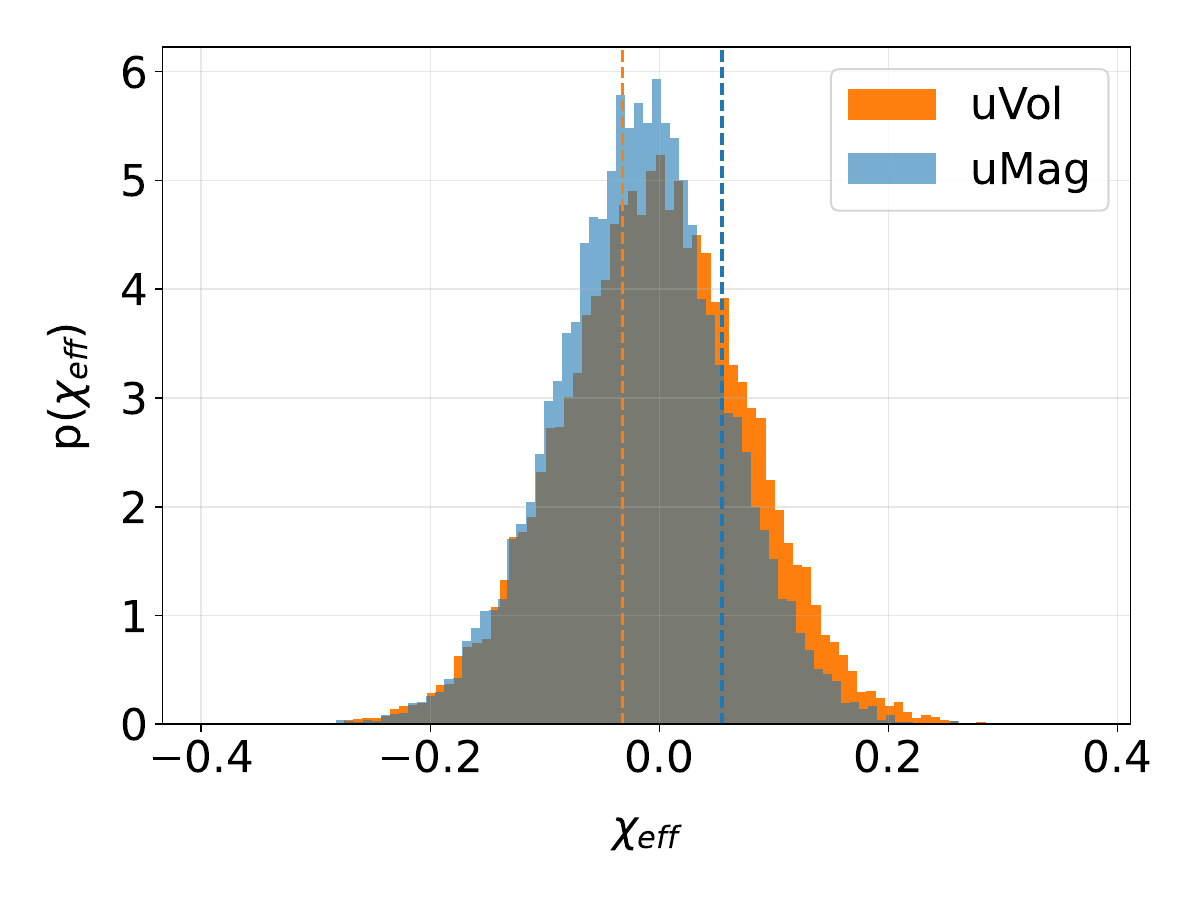}
    \includegraphics[width=0.32\linewidth]{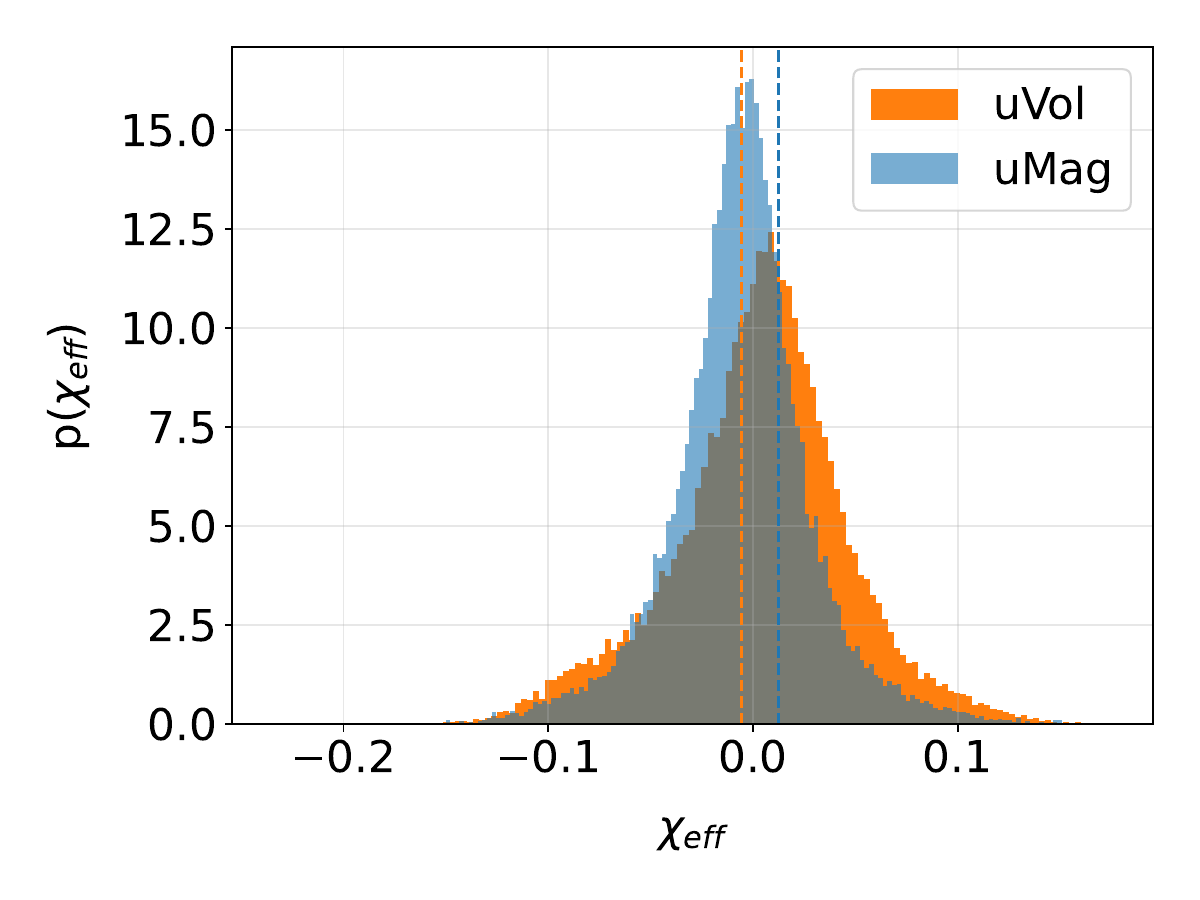}
    \includegraphics[width=0.32\linewidth]{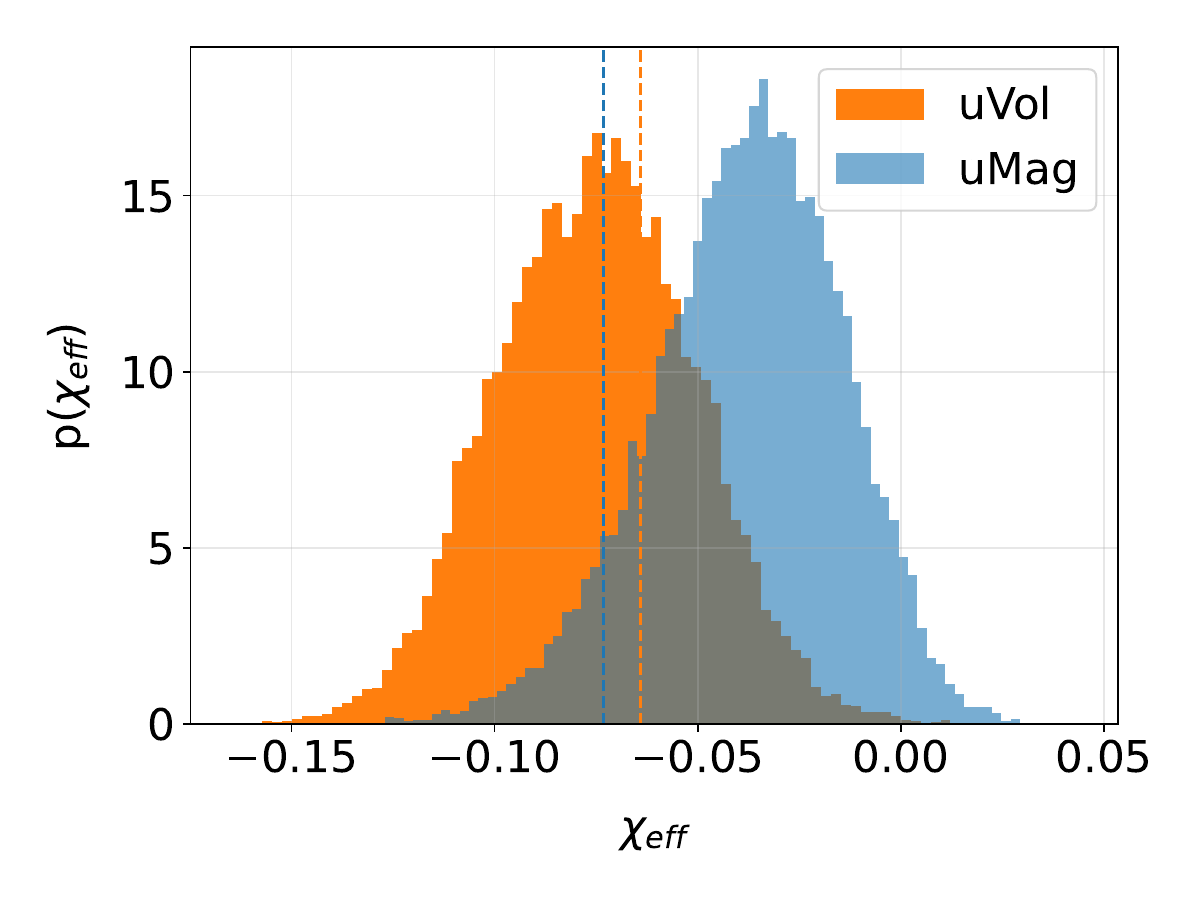}\\
    \includegraphics[width=0.32\linewidth]{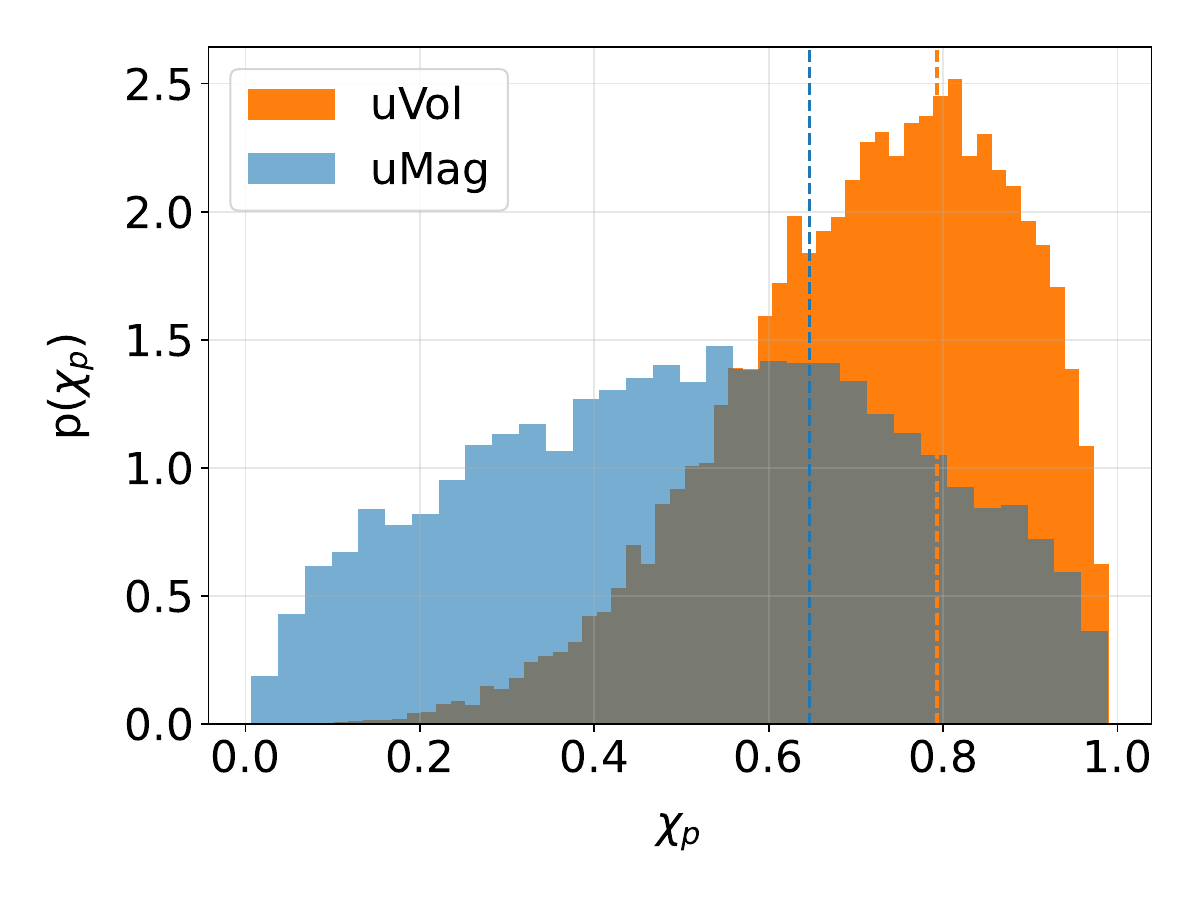}
    \includegraphics[width=0.32\linewidth]{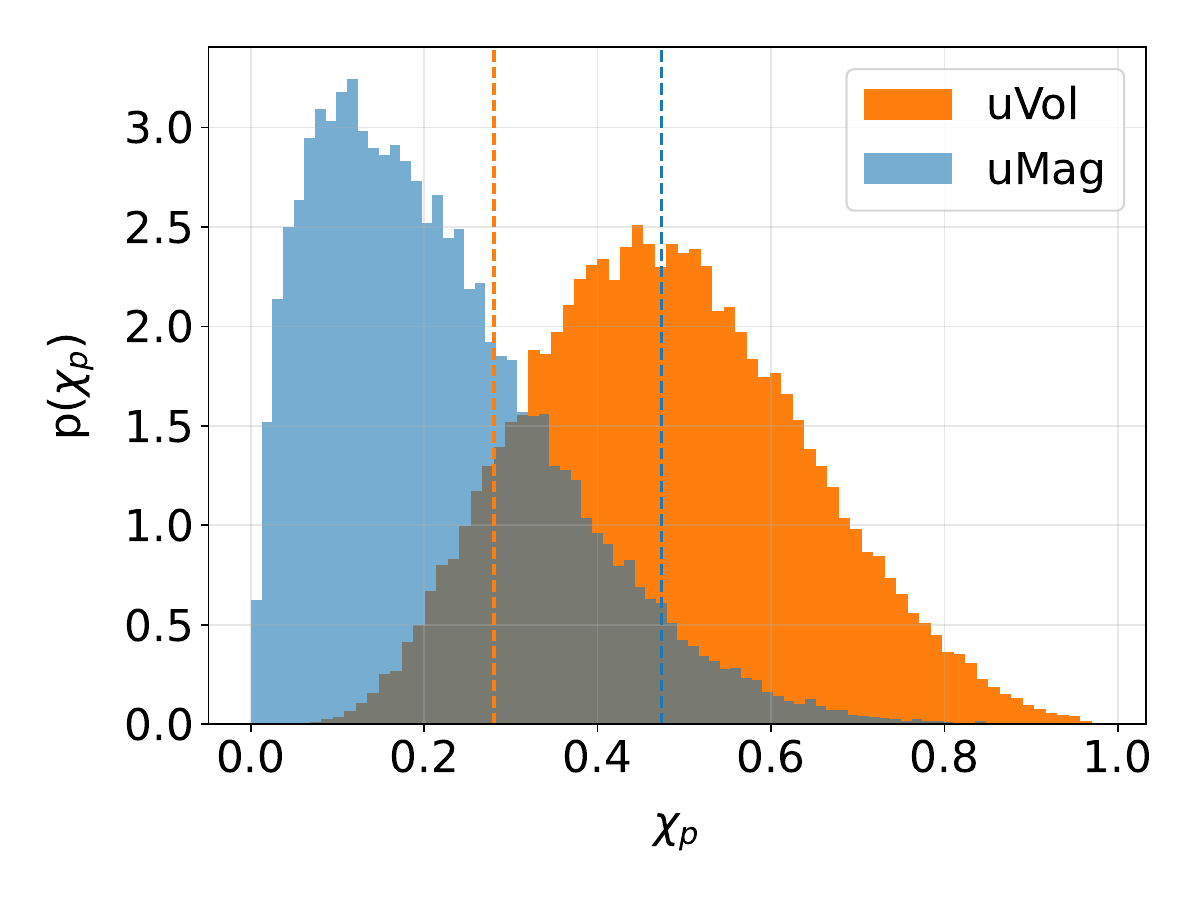}
    \includegraphics[width=0.32\linewidth]{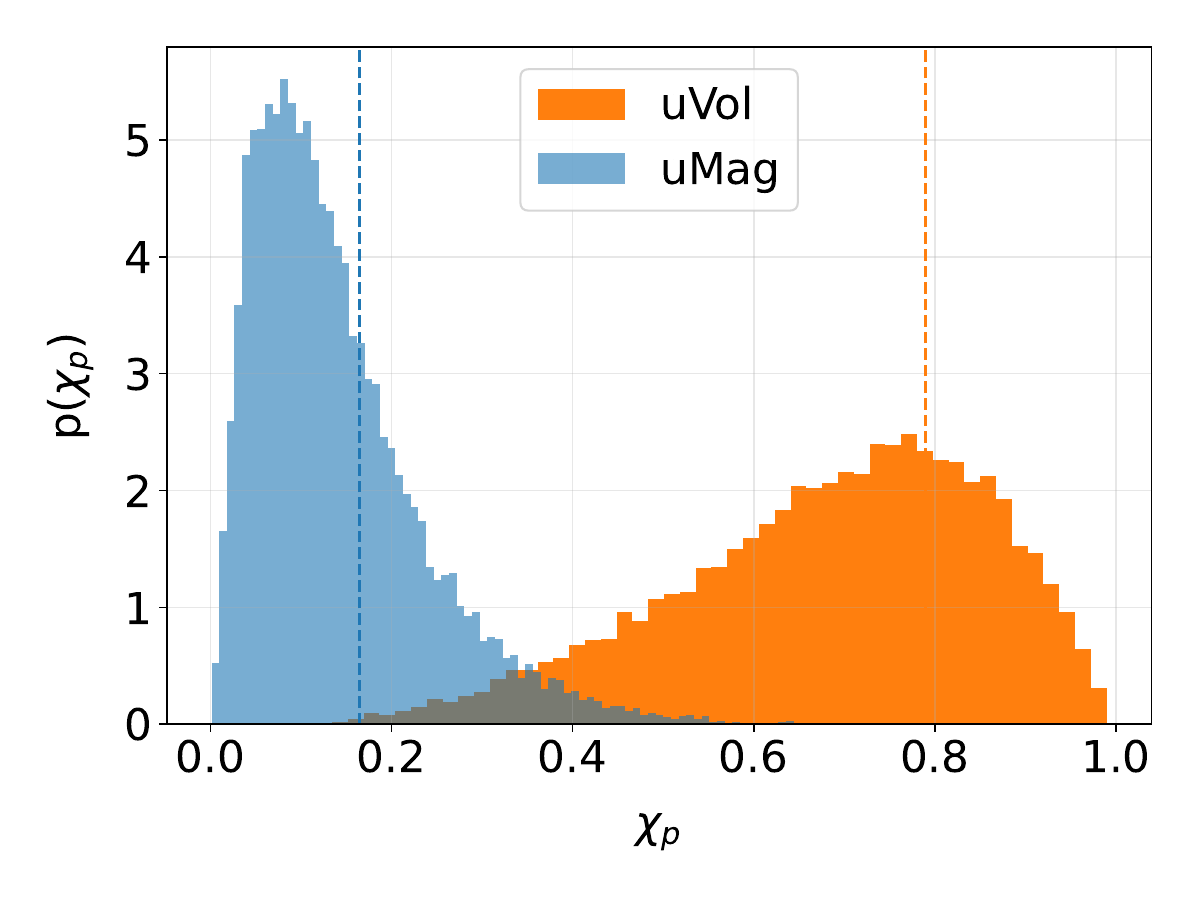}\\
    \includegraphics[width=0.32\linewidth]{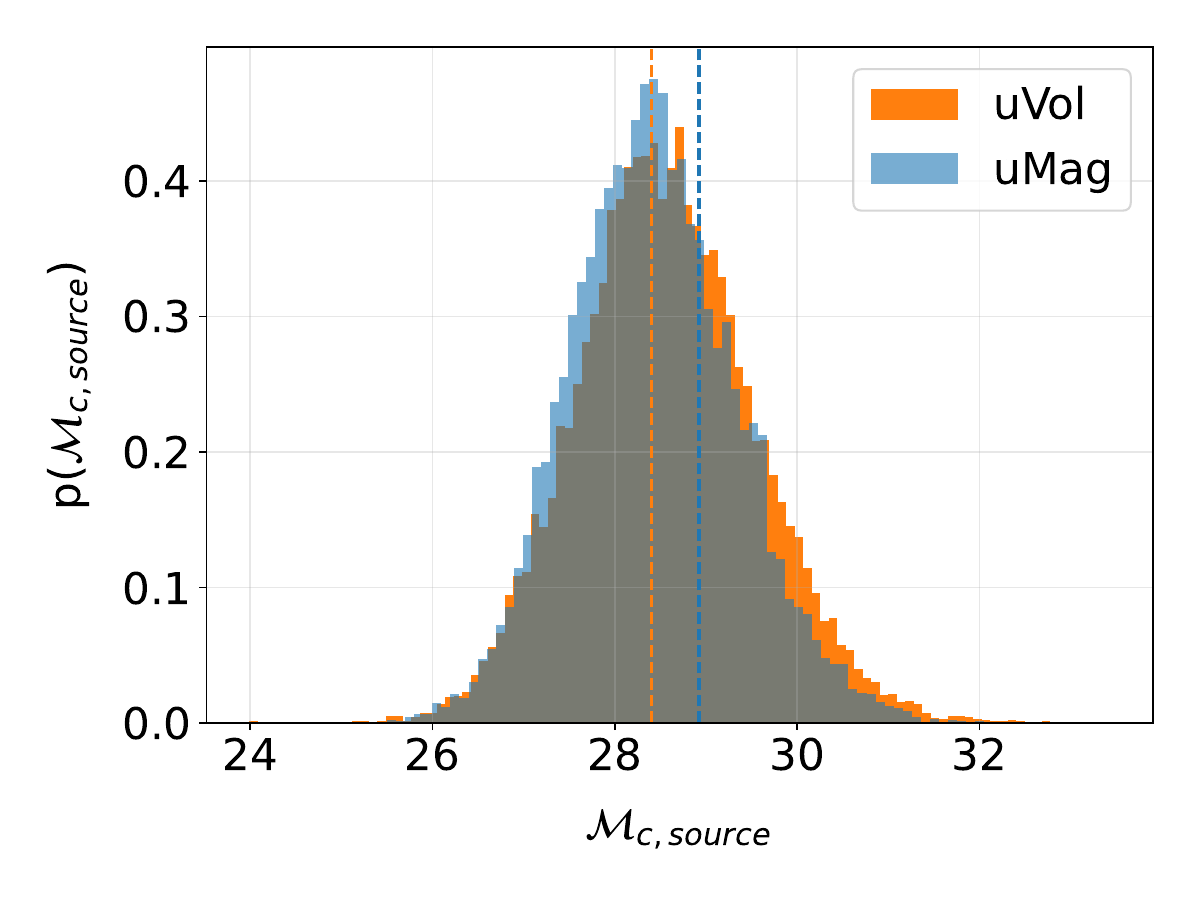}
    \includegraphics[width=0.32\linewidth]{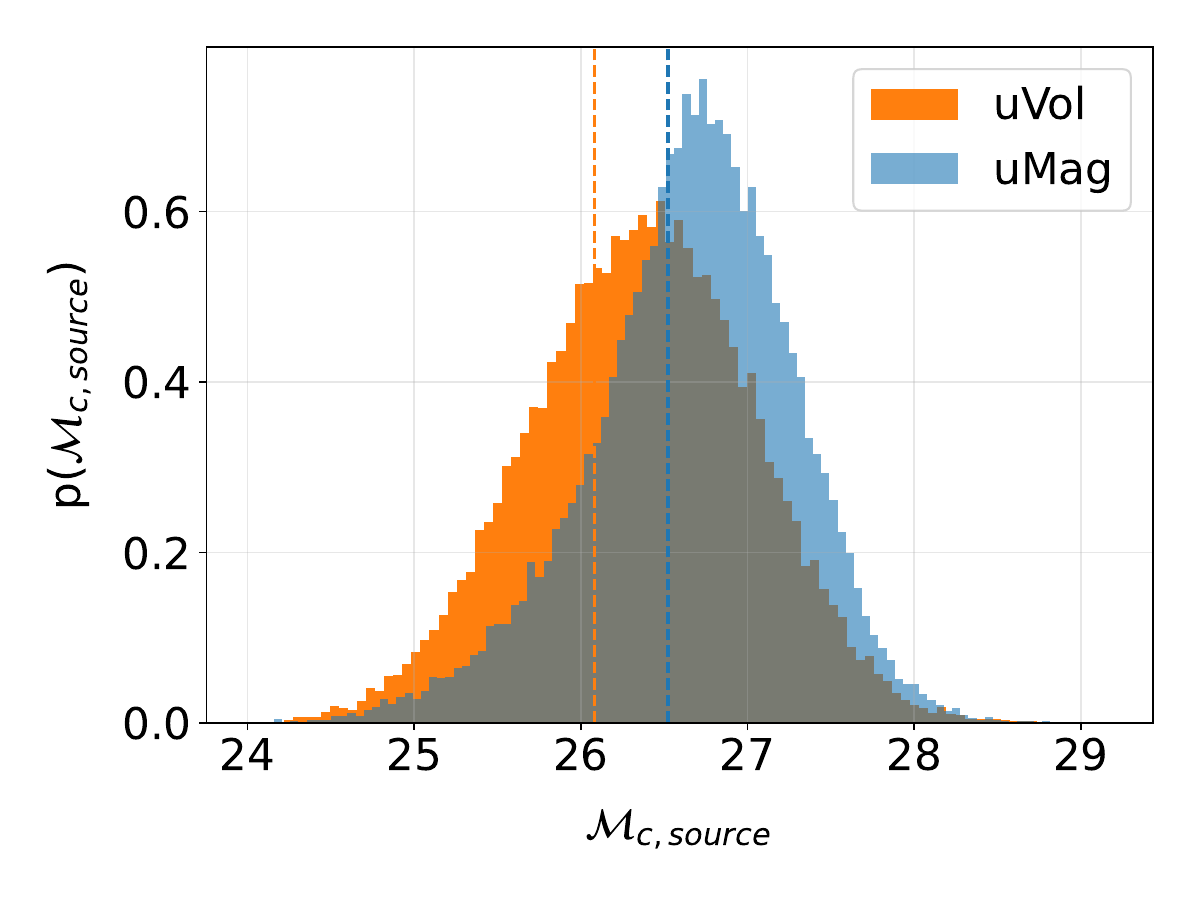}
    \includegraphics[width=0.32\linewidth]{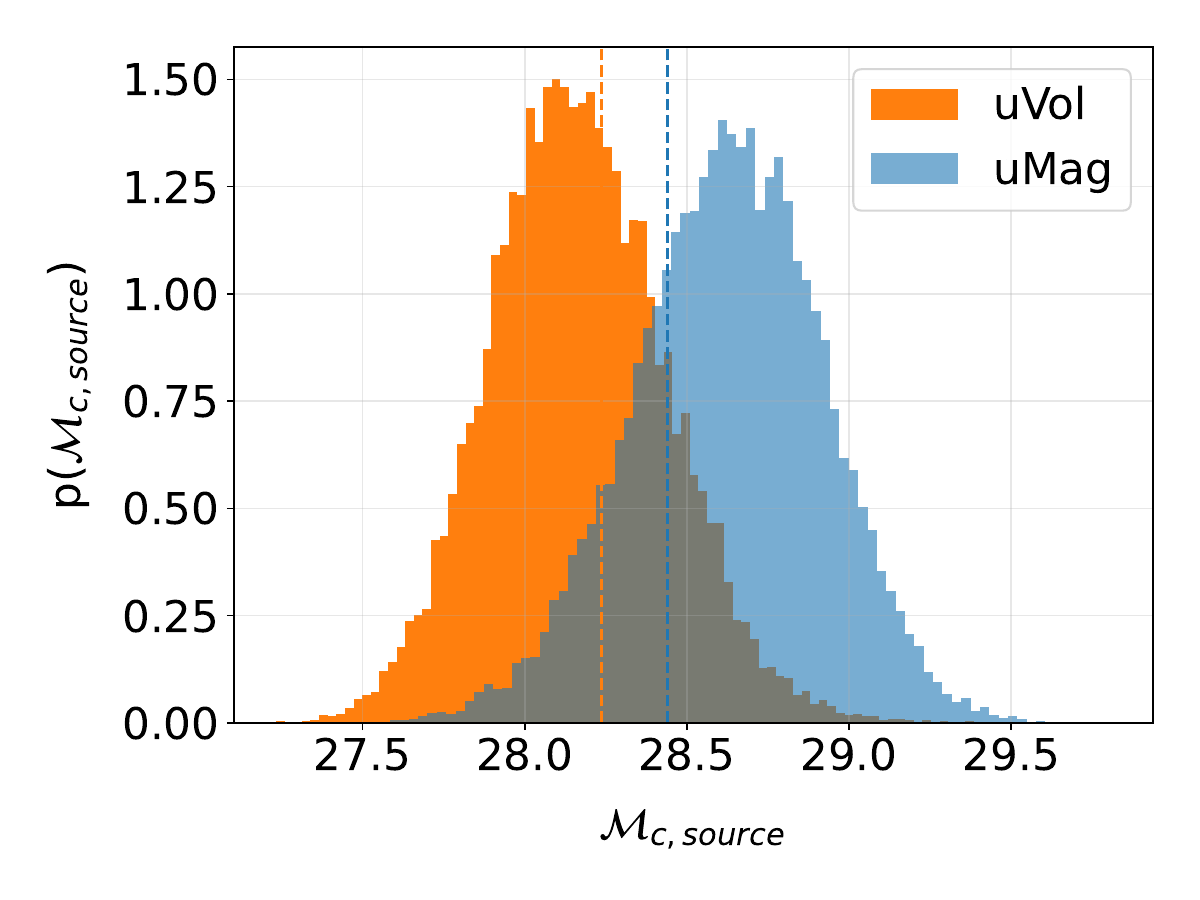} \\
    \includegraphics[width=0.32\linewidth]{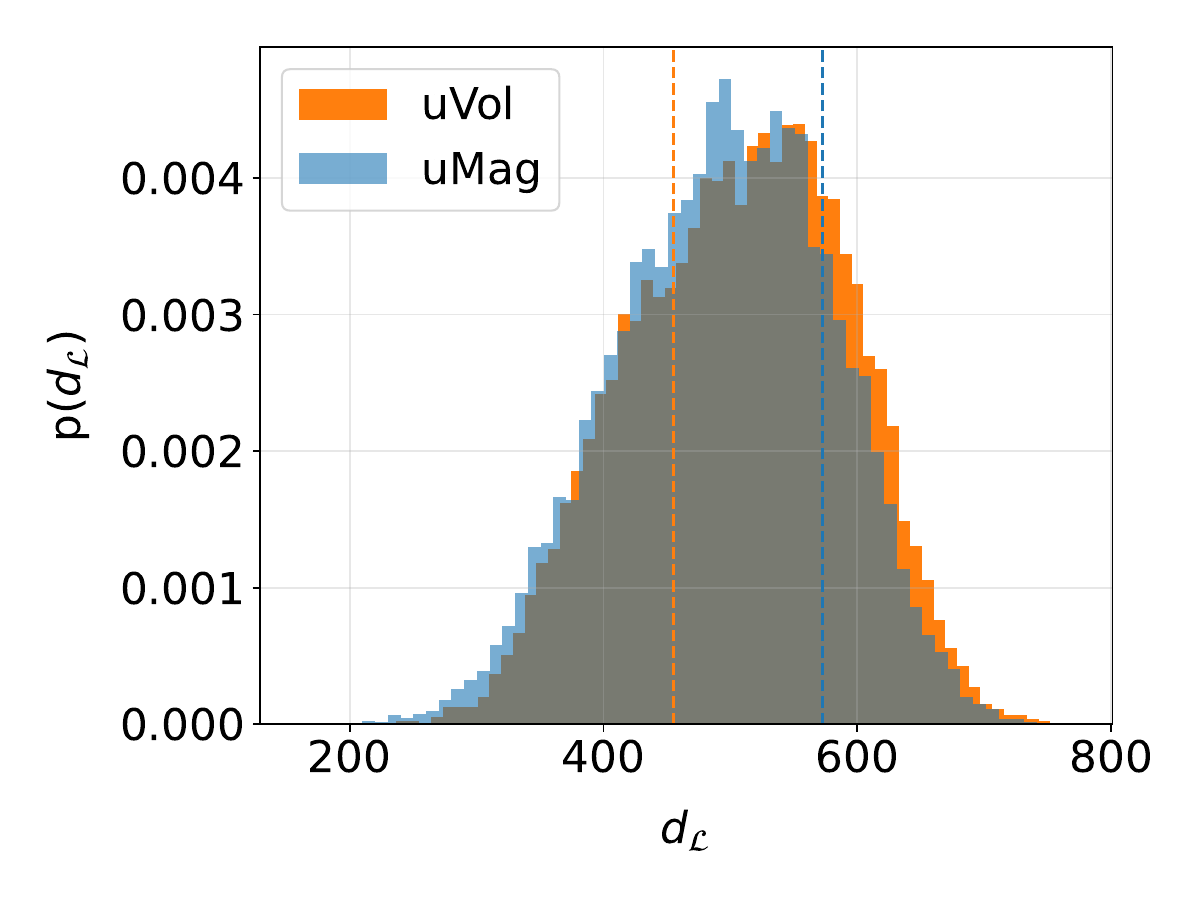}
    \includegraphics[width=0.32\linewidth]{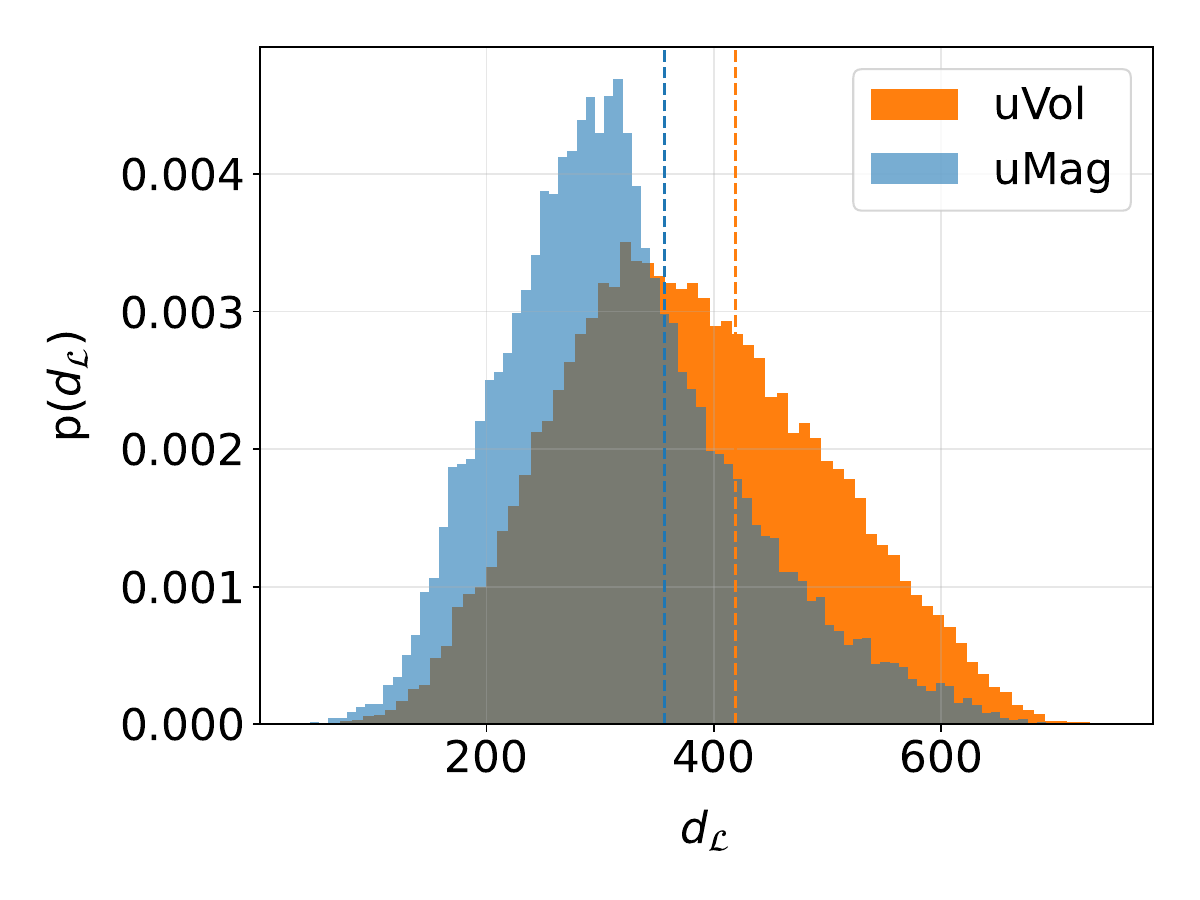}
    \includegraphics[width=0.32\linewidth]{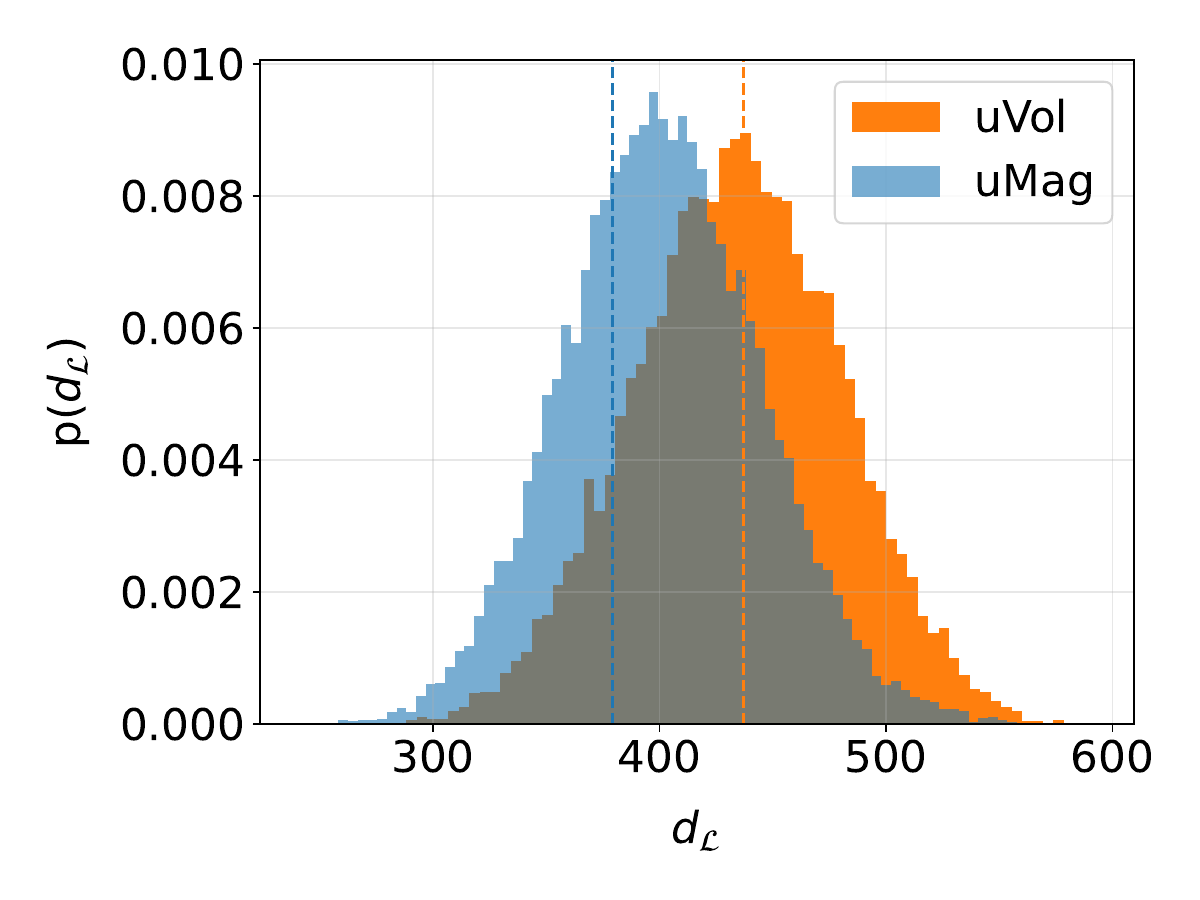} \\
    \caption{From top to bottom: the marginal distributions of network matched filter SNRs $\rho_{mf}$,  the effective spin parameter $\chi_{eff}$,  the spin-precession parameter $\chi_p$, the source frame chirp mass $\mathcal{M}_{c, source}$, and the luminosity distance $d_{\mathcal{L}}$ for GW150914 (left), GW230814 (centre) and GW250114 (right). The uniform-in-volume prior consistently yields marginally higher values of $\rho_{mf}$.}
    \label{fig:pe}
\end{figure*}
Analyzing the exceptional events with the uniform-in-volume priors not only affects the estimates of the in-plane components of spins but also marginally affects other parameters. In Fig.~\ref{fig:pe}, we plot the marginal distributions of some parameters with the uniform-in-magnitude (blue) and uniform-in-volume (orange) priors. Also plotted is the corresponding value of the maximum likelihood sample as a dashed vertical line. One can see that the network matched filter SNR is marginally higher for the latter for all three events. The effective spin distributions are not entirely different between the two priors, but are noticeable for GW250114, shifting to more anti-aligned values. The marginal distributions for the spin-precession parameter are plotted in row 3. Significant differences exist between the posteriors, and the uVol prior predicts a higher value for $\chi_p$, with the maximum likelihood value also being higher for GW150914 and GW250114, but a lower value for GW230814. For GW150914, as there is not sufficient information in the likelihood, the posteriors for $\chi_p$ recover the priors in the right panel of Fig.~\ref{fig:eff_priors}.

\paragraph{Error in $hat{p}_R$}
In this section, we compute the error in the estimator for the mean of the posterior mass in a region $R$ directly from the importance samples.

$\hat{p}_R$ can be viewed as the ratio of two random variables $\hat{A}/\hat{B}$ where $\hat{A} = \sum_j w_j I(\theta_j)$ and $\hat{B} = \sum_j w_j$, where $w_j$ are the un-normalized importance weights in the nested sampling procedure, and $I_j = I(\theta_j)$ is the mask for the region $R$, attaining one inside and zero otherwise. Now, with each run, the estimators yield a different value:
\begin{eqnarray}
    \hat{p}_R = \dfrac{\Bar{A} + \delta A}{\Bar{B} + \delta B}
\end{eqnarray}
Where the overbar denotes the mean across many runs. Assuming the deviations $\delta$ from the mean are small, the denominator can be expanded in series, leading to the first-order approximation:
\begin{align}
     \hat{p}_R &=  \dfrac{ \Bar{A}}{\Bar{B}} \left(1  + \dfrac{\delta A}{\bar{A}}\right) \left( 1 - \dfrac{\delta B}{\bar{B}} \right) + O(\delta^2)\\
      \hat{p}_R - \Bar{p}_R   &=  \dfrac{1}{\bar{B}} (\delta A - \bar{p}_R \delta B)
\end{align}
Substituting back $A$ and $B$ in terms of the un-normalized weights, one finds:
\begin{equation}
    \hat{p}_R - \bar{p}_R \approx \dfrac{1}{\sum_j w_j} \sum_i w_i (I_i - \bar{p}_R)
\end{equation}
Therefore an estimator for the variance of $\hat{p}_R$ is
\begin{equation}
    var(\hat{p}_R) = \sum_i W_i^2 (I_i - \hat{p}_R)^2 
\end{equation}
Where $W_i = \dfrac{w_i}{\sum_j w_j}$ are the normalized importance weights, and are assumed to not change at first order from one run to another.

\end{document}